\documentclass[aps,twoside,twocolumn]{revtex4}

\usepackage{amsmath}
\usepackage{amsfonts}
\usepackage{amssymb}
\usepackage{epsfig}

\def\jtitle#1{#1}   % use this line to include titles of journal articles

\newtheorem{theorem}{Theorem}
\newtheorem{lemma}[theorem]{Lemma}
\newtheorem{definition}[theorem]{Definition}
\newtheorem{proposition}[theorem]{Proposition}
\newtheorem{corr}[theorem]{Corollary}
\def\proof{\par\vskip12pt\noindent{\it Proof}}
\def\qed{\leavevmode\unskip\penalty9999 \hbox{}\nobreak\hfill
     \quad\hbox{\leavevmode  \hbox to.77778em{%
               \hfil\vrule   \vbox to.675em%
               {\hrule width.6em\vfil\hrule}\vrule\hfil}}}

\def\AA{{\cal A}}\def\BB{{\cal B}}\def\DD{{\cal D}}
\def\HH{{\cal H}}
\def\MM{{\cal M}}\def\ZZ{{\cal Z}}
\def\idty{\hbox{\rm\openone}}
\def\tr{{\rm tr}}

\def\norm#1{\left\Vert{#1}\right\Vert}
\def\Cx{{\mathbb{C}}}\def\Ir{\hbox{$\mathbb{Z}$}}
\def\ket#1{\vert#1\rangle}
\def\bra#1{\langle#1\vert}
\def\ketbra#1#2{\vert#1\rangle\langle#2\vert}

\def\Nei{{\cal N}}% Neighbourhood scheme
\def\loclat{{\cal L}} % quotient lattice for local rules
\def\qone{{\bf1}}
\def\dtwos{d^{\,2^{\scriptstyle s}}}
\def\grule{T}
\def\lrule{\grule_0}
\def\brule{\grule_\Box}

\def\spp{{\bf s}} %linear support of subspace of tensor product
\def\Spp{{\bf S}} % support algebra

\def\OO{{\mathcal O}}
\def\AO#1{{\mathfrak A}({\OO}_{#1})}
\def\spacelike{\hbox{$/\mkern-9mu\backslash$}}
\def\Field{{\mathbb F}}

\begin{document}
%* Top Matter
\title{Reversible Quantum Cellular Automata}
\author{B. Schumacher}
\email{schumacherb@kenyon.edu} \affiliation{Kenyon College, Dept.
of Physics, Gambier, Ohio 43022-9623, USA}
\author{R.~F. Werner}
\email{R.Werner@TU-BS.DE} \affiliation{Inst. Math. Phys.,
TU-Braunschweig, Mendelssohnstra{\ss}e 3,
  D-38106 Braunschweig, Germany}
\begin{abstract}
We define quantum cellular automata as infinite quantum lattice
systems with discrete time dynamics, such that the time step
commutes with lattice translations and has strictly finite
propagation speed. In contrast to earlier definitions this allows
us to give an explicit characterization of all local rules
generating such automata. The same local rules also generate the
global time step for automata with periodic boundary conditions.
Our main structure theorem asserts that any quantum cellular
automaton is structurally reversible, i.e., that it can be
obtained by applying two blockwise unitary operations in a
generalized Margolus partitioning scheme. This implies that, in
contrast to the classical case, the inverse of a nearest neighbor
quantum cellular automaton is again a nearest neighbor automaton.

We present several construction methods for quantum cellular
automata, based on unitaries commuting with their translates, on
the quantization of (arbitrary) reversible classical cellular
automata, on quantum circuits, and on Clifford transformations
with respect to a description of the single cells by finite Weyl
systems. Moreover, we indicate how quantum random walks can be
considered as special cases of cellular automata, namely by
restricting a quantum lattice gas automaton with local particle
number conservation to the single particle sector.

\end{abstract}
\maketitle

\section{Introduction}
The idea of generalizing the classical notion of cellular automata
to the quantum regime is certainly not new. Indeed, it is already
present in Feynman's famous paper \cite{Feynman} from 1982, in
which he argues that quantum computation might more powerful than
classical. However, although there have been several formal
definitions of quantum cellular automata over the years, the
theory is not in good shape at the moment, and a systematic
exploration of the general properties of such systems on the one
hand and of the potential for computational applications on the
other has hardly begun. We believe that this is partly due to
deficiencies of the existing approaches, and therefore propose a
new one, which is very natural, and only requires a few basic
assumptions: a discrete cell structure with a finite quantum
system for every cell and translation symmetry, a discrete time
step for the global system, reversibility, and finite propagation
speed.

Quantum cellular automata (``QCAs'') are of interest to several
fields. There are obvious connections to the statistical mechanics
of lattice systems, and potential applications to ultraviolet
regularization of quantum field theories. In Quantum Computer
Science they appear as one natural model of computation extending
the well-developed theory of classical cellular automata into the
quantum domain. But also the experimental side is rapidly
developing: quantum computing in optical lattices \cite{optlat}
and arrays of microtraps \cite{mictrap} are among the most
promising candidates for the first quantum computer that does
useful computations \cite{QdCA}. It is typical for such systems
that the addressing of individual cells is much harder than a
change of external parameters affecting all cells in the same way
\cite{Benjamin}. But this is just the theoretical description of a
cellular automaton. Possible tasks, which have the same built-in
translation invariance are simulations of solid state models.
Typically classical simulations run into problems already for
moderate systems sizes, precisely because of the dimension and
complexity explosion which Feynman noted, and for which he
proposed quantum computation as a cure. The theory presented in
this paper can be considered as providing the first elements of an
assembly language for such simulations.

The problems which have plagued previous attempts to define QCAs
begin with the definition of a system of infinitely many cells.
Consider the simplest operation such an automaton should be able
to perform: applying the same unitary transformation separately to
each cell. This would involve multiplying infinitely many phases,
so there is really no well-defined unitary operator describing the
global state change. Therefore, the quantization approach ``just
make the transition function unitary'' does not work very well.
Similarly, the notion of state vectors as amplitude assignments to
uncountably many classical configurations is ill-defined. But this
would be a candidate for the ``configurations'' of a QCA, which
causes problems for a definition of QCAs in terms of
configurations and their transformations. Various approaches
\cite{Watrous,vanDam,Gruska,santa,Feynman} will be described and
commented in Section~\ref{sec:others}. A constructive method to
obtain QCAs, which is common to most of these approaches
(including ours) is partitioning the system into blocks of cells,
applying blockwise unitary transformations, and possibly iterating
such operations. Model studies based on such constructions (e.g.,
\cite{Brennen}) therefore produce results independently of the
definition problems.

In order to arrive at a satisfactory notion of QCAs, it is helpful
to draw on ideas from a discipline, which has been dealing with
infinite arrays of simple quantum systems for a long time, i.e.,
the statistical mechanics of quantum spin systems. Infinite
systems have been considered particularly in the algebraic
approach to such systems \cite{BraRo}. The basic idea is to focus
on the observables rather than the states, i.e., to work in the
Heisenberg picture rather than the Schr\"odinger picture
\cite{Paschen}. The main advantage is that in contrast to
localized states, it does make sense speak of local observables
\cite{Haag}, i.e., observables requiring a measurement only of a
finite collection of cells. The global transition rule of a
cellular automaton is then a transformation $T$ on the observable
algebra of the infinite system. As always in the Heisenberg
picture, the interpretation of such a transformation is that
`preparing a state, running the automaton for one step, and then
measuring the observable $A$' gives exactly the same expectations
as preparing the same state and measuring $T(A)$. As always, $T$
must be completely positive and satisfy $T(\idty)=\idty$. But more
importantly we can state the crucial localization property of
QCAs: When $A$ is localized on a region $\Lambda$ of the lattice,
then $T(A)$ should be localized in
$\Lambda+\Nei=\{x+n|\;x\in\Lambda,\ n\in\Nei\}$, where $\Nei$ is
the {\it neighborhood scheme} of the QCA.

To our knowledge, this view of QCAs was first used in \cite{RiWe},
where the approach to equilibrium in a QCA with irreversible local
rules (based on a partitioning scheme) was investigated. The
general picture of QCAs remained unsatisfactory, however, because
the partitioning scheme seemed a rather special way of
constructing a QCA. For a satisfactory theory of QCAs we demand
that there should be a direct connection between the {\it global
transition rule} $T$ and the {\it local transition rule}: If we
know the global transition rule, we should be able to extract
immediately the local rule in a unique way, and conversely, from
the local rule we should be able to synthesize the global rule.
The class of global rules should have an axiomatic specification,
the most important of which would be the existence of a finite
neighborhood scheme in the above sense. On the other hand, for the
local transition rules we would prefer a {\it constructive
characterization}. That is, there should be a procedure for
obtaining all local rules leading to global rules with the
specified properties, in which all choices are clearly
parameterized.

The partitioning QCAs of \cite{RiWe} failed to meet these
requirements, because they provided a construction, but no
axiomatic characterization of the global rules obtained in this
way. In particular, it remained unclear whether two steps of such
an automaton could be considered as a single step of an automaton
with enlarged neighborhood scheme. The idea enabling the present
paper was that all these difficulties vanish if we restrict to the
class of {\it reversible} QCAs. The axiomatic characterization is
extremely simple: In addition to the above locality condition we
assume that the global rule must have an inverse, which is again
an admissible quantum channel. This is equivalent to saying that
$T$ must be an automorphism of the observable algebra. Then the
local rule is simply the restriction of this automorphism to the
algebra of a single cell. Conversely, since every observable can
be obtained as a linear combination of products of single-cell
observables, the local rule determines the global automorphism.
This allows us to subsume all the known constructions of QCAs, but
also to prove a general structure theorem: every reversible QCA is
structurally reversible, i.e., we can write the local rule in a
partitioning scheme involving two unitary matrices, which makes it
apparent how the global rule can be unitarily implemented on
arbitrarily large regions, and how to obtain the local rule of the
inverse.

A further bonus from our proof of the structure theorem is that it
does not actually require the global rule to be an automorphism:
it works under the prima facie much weaker assumption that the
global rule is a homomorphism (and not necessarily onto). Then
invertibility follows (see Corollary~\ref{cor:invert} below).
Therefore, invertibility was not included in Definition~1, which
makes it much easier to verify whether a proposed rule is indeed a
QCA.

The structural invertibility was an open problem in the theory of
classical reversible cellular automata in higher dimensional
lattices until recently \cite{Kari1}. Hence, since our proof of
the quantum result is rather simple, it appears that some proofs
in the classical domain can be simplified by going quantum.

Our paper is organized as follows: In Section~\ref{sec:def}, we
begin with the axiomatic definition of QCAs in the sense described
above. Its counterpart, the constructive description is given in
the form of a collection of basic constructions and examples in
Section~\ref{sec:basicon}. It turns out that one of these
constructions, based on partitioning, is already sufficient to
obtain all QCAs in the sense of our definition. This rather
surprising result is stated and proved in Section~\ref{sec:struc}.
The ideas of the proof also allows us to give an explicit
parameterization of the simplest class of QCAs: nearest neighbor
automata in one dimension with one qubit per cell. As mentioned in
the introduction, the current literature on the subject is mostly
based on a definition we do not find satisfactory. We discuss
these, and some further related definitions, in more detail in
Section~\ref{sec:others}. Finally, in an appendix we provide some
mathematical background on finite dimensional C*-algebras, which
play a key role in the proof of Theorem~\ref{mainthm}.

\section{Definition of QCAs}\label{sec:def}
We consider an infinite cubic array of cells, labelled by integer
vectors $x\in\Ir^s$, where $s\geq1$ is the spatial dimension of
the lattice \cite{otherlat}. Each cell contains a $d$-level
quantum system with the same finite $d\geq2$. That is to say, with
each cell $x\in\Ir^s$, we associate the observable algebra $\AA_x$
of the cell, and each of these algebras is an isomorphic copy of
the algebra of complex $d\times d$-matrices.  When
$\Lambda\subset\Ir^s$ is a finite subset, we denote by
$\AA(\Lambda)$ the algebra of observables belonging to all cells
in $\Lambda$, i.e., the tensor product
$\bigotimes_{x\in\Lambda}\AA_x$. By tensoring with unit operators
on $\Lambda_2\setminus\Lambda_1$ we consider $\AA(\Lambda_1)$ as a
subalgebra of $\AA(\Lambda_2)$, whenever
$\Lambda_1\subset\Lambda_2$. In this way the product $A_1A_2$ of
$A_i\in\AA(\Lambda_i)$ becomes a well-defined element of
$\AA(\Lambda_1\cup\Lambda_2)$. Moreover, tensoring with the
identity does not change the norm, so we get a normed algebra of
{\it local observables}, whose completion is called the {\it
quasi-local algebra}\cite{BraRo}, and will be denoted by
$\AA(\Ir^s)$. Similarly, for other infinite subsets
$\Lambda\subset\Ir^s$ we define $\AA(\Lambda)$ as the closure of
the union of all $\AA(\Lambda')$ with $\Lambda'\subset \Lambda$
finite.

When $x\in\Ir^s$ is a {\it lattice translation}, we denote by
$\tau_x$ the isomorphism from each $\AA_y$ to $\AA_{x+y}$, and its
extensions from $\AA(\Lambda)\to\AA(\Lambda+x)$, by shifting every
site. Here we have used the notation
$\Lambda+x=\{y+x|y\in\Lambda\}$ for shifted lattice subsets, which
we also extend to
$\Lambda_1+\Lambda_2=\{x_1+x_2|x_i\in\Lambda_i\}$.

A {\it state} $\omega$ of the spin system is a linear functional
on $\AA(\Ir^s)$, which is positive in the sense that
$\omega(X^*X)\geq0$ and normalized as $\omega(\idty)=1$.
Equivalently, a state is given by a family $\omega_\Lambda$ of
density operators on $(\Cx^d)^{\otimes\Lambda}$ (for each finite
$\Lambda$), such that
 $\omega(X)=\tr(\omega_\Lambda X)$ for $X\in\AA(\Lambda)$. The
local density matrices have to satisfy the consistency condition
that, for $\Lambda_1\subset\Lambda_2$, $\omega_{\Lambda_1}$ is
obtained from $\omega_{\Lambda_2}$ by tracing out all tensor
factors in $\Lambda_2\setminus\Lambda_1$. Note that a state does
not correspond to a configuration of a classical automaton, but
rather to a probability distribution over global configurations.

\begin{definition}
 A {\bf Quantum Cellular Automaton}
 with neighborhood scheme $\Nei\subset\Ir^s$
 is an homomorphism $\grule:\AA(\Ir^s)\to\AA(\Ir^s)$ of the quasi-local
algebra, which commutes with lattice translations, and satisfies
the locality condition
 $\grule(\AA(\Lambda))\subset\AA(\Lambda+\Nei)$ for every finite
set $\Lambda\subset\Ir^s$.
 The local {\bf transition rule} of a cellular automaton is the
homomorphism $\lrule:\AA_0\to\AA(\Nei)$. \end{definition}

Note that a unitary operator for the time evolution is not necessary in
this formulation. Instead we have replaced it by its action on
observables. Of course, the time step has to be read in the Heisenberg
picture. That is,  measuring some local observable $A\in\AA(\Lambda)$ at
time $t+1$ is equivalent to measuring the observable $\grule(A)$ at time
$t$. The transition rule thus describes this backwards calculation for a
single cell. $\lrule(\AA_0)$ is some isomorphic copy of the one-cell
algebra embedded in a possibly quite complicated way into the algebra of
neighboring cells. The relationship between the global evolution and the
one-site transition rule is as simple as it should be:

\begin{lemma}\label{lem:localrule}{\ }\newline
 (1) The global homomorphism $\grule$ is uniquely determined by the
local transition rule $\lrule$. \newline
 (2) A homomorphism $\lrule:\AA_0\to\AA(\Nei)$ is the transition
rule of a cellular automaton if and only if for all $x\in\Ir^s$ such that
$\Nei\cap(\Nei+x)\neq\emptyset$ the algebras $\lrule(\AA_0)$ and
$\tau_x\bigl(\lrule(\AA_0)\bigr)$ commute elementwise. \end{lemma}

\proof:\quad  By translation invariance the action of
$\grule_x:\AA_x\to\AA(\Nei+x)$ is determined as
$\grule_x(A_x)=\tau_x\lrule\tau_{-x}(A_x)$ on all other cells.
Moreover, because $\grule$ is a homomorphism, the extension to any
local algebra is also fixed. Explicitly, consider a product
$\bigotimes_{x\in\Lambda}A_x=\prod_{x\in\Lambda}A_x$ of one-site
operators. This equation just expresses our identification of the
one site algebras $\AA_x$ with subalgebras of $\AA(\Lambda)$ by
tensoring with unit operators. Then the homomorphism property of
the global evolution requires that
\begin{equation}\label{alfaglobal}
  \grule\left(\bigotimes_{x\in\Lambda} A_x\right)
    =\prod_{x\in\Lambda} \grule_x(A_x)\;.
\end{equation}
 Note that the product on the right hand side cannot be replaced
by a tensor product, because the factors have overlapping
localization regions $x+\Nei$. Moreover, the argument of $\grule$
is a product of commuting factors, hence so is the right hand
side. Hence the the commutativity condition (2) is necessary.

Conversely, if the factors $\grule_x(A_x)$ commute, their product
is unambiguously defined. Since every local observable can be
expressed as a linear combination of tensor products,
Eq.~(\ref{alfaglobal}) defines a homomorphism on the local
algebra, as required.  This shows the converse of (2), and since
we have given an explicit formula of $\grule$ in terms of $\lrule$
it also shows (1). \qed

It is clear that the commutation condition of the Lemma can be
expressed as a finite set of equations, and can therefore be
verified effectively. Since only a small portion of the lattice is
needed in this verification, the same steps are needed to check
local transition rules fore QCAs on graphs which locally look like
$\Ir^s$. Such graphs can be seen as integer lattices with {\it
periodic boundary conditions}. The only condition we will need to
impose is that the periods for the boundary condition are not too
small compared with the size of neighborhood scheme $\Nei$ (a
condition called ``regularity'' below). No algebraic conditions on
the homomorphism $\lrule$ are needed. Therefore we can turn the
construction around, and immediately get a QCA on the infinite
lattice from a QCA with finitely many cells. Since for finitely
many cells a homomorphism is always just implemented by a unitary
matrix, this allows us to define QCAs even for the infinite
system, just by specifying unitary matrices with suitable
properties.

\begin{figure}[htb]
\epsfxsize=5cm \epsffile{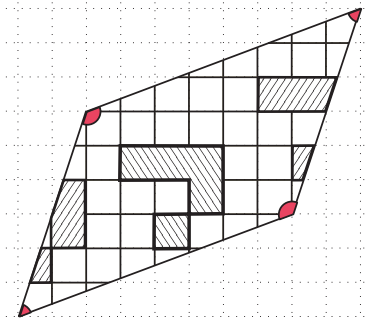}
 \caption{\label{figlattis}{\it Periodic boundary conditions.
 The parallelogram represents the given finite lattice $\loclat$,
 where cells cut by opposite boundaries must be suitably joined.
 The marks on the corners join up to a full circle.
 Two translates of the same neighbourhood are shown.
 It is regular, because the geometry of intersections is
 the same as on the infinite plane.  }}
\end{figure}

Let us describe the periodic boundary QCAs more precisely and see
what conditions are needed for the neighborhoods. The cells of a
system with periodic boundary conditions arise from the cells in
$\Ir^s$ by identifying certain cells, namely all those differing
by a vector $\gamma$ in some subgroup $\Gamma\subset\Ir^s$. The
set of cells is thus identified with the quotient
$\loclat=\Ir^s/\Gamma$. For each point in $x\in\loclat$, i.e.,
each equivalence class $x=x_0+\Gamma$ of identified cells, only
one observable algebra is given. The sites $\loclat$ can be
represented in a so called fundamental domain of $\Gamma$ like the
parallelogram in Fig.~\ref{figlattis}. But there seems to be no
way to draw this nicely as a set of square cells.

\begin{figure}[htb]
\epsfxsize=2.5cm \epsffile{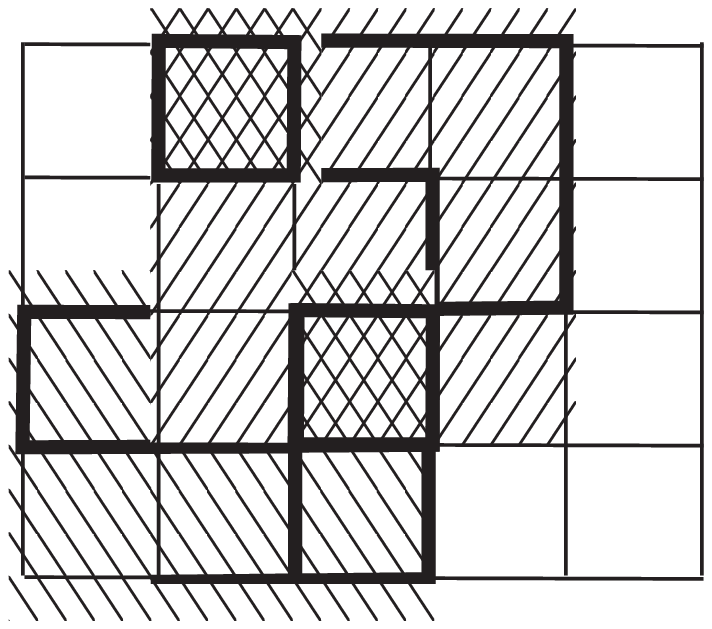}
 \caption{\label{figirreg}{\it The same neighborhood scheme
 is not regular on a 5$\times$4 torus: Again we have two translates,
 but their intersection (cross-hatched) cannot be realized on the
 infinite square lattice as intersection of two such neighborhoods.  }}
\end{figure}

For points $x=(x_0+\Gamma)\in\loclat$, i.e., we can define the
translation $x+n=x_0+n+\Gamma$. Consider now a neighborhood scheme
$\Nei\subset\Ir^s$. The neighborhood of $x\in\loclat$ is then
$x+\Nei\subset\loclat$. Note that as far as the model with
periodic boundary conditions is concerned we could change each
$n\in\Nei$ by a lattice vector in $\Gamma$ without changing the
neighborhood of any point. But since we are interested in the
connection with the infinite model, we do not admit this
ambiguity. The neighborhood scheme is called {\it regular} for the
given periodic structure given by $\loclat$, if the equations we
have to check for the commutation rule of a cellular automaton are
the same in both cases. Since these equations depend only on the
intersections between translates of $\Nei$ we only need to make
sure that the geometry of intersections is the same. So suppose
that neighborhoods on $\loclat$ intersect, say
$(x+\Nei)\cap(y+\Nei)\neq\emptyset$. This means that there is a
translation $m\in\Ir^s$ such that $x+m=y$ and also
$\Nei\cap(m+\Nei)\neq\emptyset$. Clearly, the first condition
determines $m$ up to a lattice translation $\Gamma$, and the
second is an intersection condition on the infinite lattice.
Obviously, $x+(\Nei\cap(\Nei+m))\subset(x+\Nei)\cap(y+\Nei)$. But
the inclusion could be strict if there is more than one $m$ with
the required properties. This is precisely the case regularity
must exclude. To summarize, we call a neighborhood scheme
$\Nei\subset\Ir^s$ {\it regular} for a subgroup
$\Gamma\subset\Ir^s$, if $\Nei\cap(m+\Nei)\neq\emptyset$ and
$n\in\Gamma$, $n\neq0$, imply that $\Nei\cap(m+n+\Nei)=\emptyset$.
A more compact equivalent form is
$(\Nei+\Nei-\Nei-\Nei)\cap\Gamma=\{0\}$. An example of a regular
neighborhood is given in Figure~\ref{figlattis}. That the same
neighborhood becomes non-regular for a smaller lattice is shown in
Fig.~\ref{figirreg}.

 Then checking condition (2) of the Lemma for the QCA
on $\loclat$ and for the QCA on $\Ir^s$ are exactly equivalent. We
call this useful principle the {\it Wrapping Lemma:}

\begin{lemma} The QCA transition rules on a finite lattice
$\loclat$ with respect to a regular neighborhood scheme $\Nei$ are
in one-to-one correspondence with the transition rules for QCAs on
$\Ir^s$ with the same neighborhood scheme. \end{lemma}

\par

\section{Basic Construction Methods}\label{sec:basicon}
\subsection{Commuting Unitaries and Phases }\label{sec:communitary}
Consider a unitary operator $U_0$ in some local algebra
$\AA(\widetilde\Nei)$, which commutes with all its translates
$U_x=\tau_x(U_0)$, up to a phase. That is, we require that there
are complex numbers $\zeta_x$ with $|\zeta_x|=1$ such that
$U_0\tau_x(U_0)=\zeta_x\tau_x(U_0)U_0$ for
$(\widetilde\Nei+x)\cap\widetilde\Nei\neq\emptyset$. This is
equivalent to
\begin{equation}\label{localprojective}
  U_xU_y=\zeta_{y-x}\;U_yU_x
\end{equation}
We can then formally define a unitary operator
\begin{equation}\label{infUnitary}
    \mbox{``\ }U=\prod_{{x\in\Ir^s}} U_x
    \mbox{\ ''.}
\end{equation}
 Here the the scare quotes indicate that there is no way this
infinite product can be made sense of. However, we can define
instead the action of this ``operator'' on local observables. To
this end, note that the actions $A\mapsto U_x^*AU_x$ commute for
different $x$. Moreover, if $x+\Nei$ does not intersect the
localization region of $A$, this action is the identity.
Therefore, in the infinite product of these operations only a
finite set is {\it not} the identity, and their product defines an
automorphism $\grule$. More formally, we have
\begin{equation}\label{commutealpha}
  \grule(A)=\lim_{\Lambda\nearrow\Ir^s} U_\Lambda^*AU_\Lambda\;,
\end{equation}
 where $A\in\AA(\Ir^s)$, and
 $U_\Lambda=\prod_{{x\in\Lambda}}U_x$.
The limit is over any sequence of finite sets, eventually
absorbing all lattice points, and if $A$ is localized in a finite
region, the limit is actually constant for sufficiently large
$\Lambda$.

 The local transition rule is found by
applying $\grule$ to a single cell. This gives the neighborhood scheme
\begin{equation}\label{neiCommute}
  \Nei=\bigcup_x\{\widetilde\Nei+x| 0\in\widetilde\Nei+x\}
      =\widetilde\Nei-\widetilde\Nei\;.
\end{equation}

We mention three special cases of this construction:
\begin{itemize}
\item When $\widetilde\Nei=\Nei=\{0\}$, we have to choose a unitary operator
$U_0\in\AA_0$ acting on a single cell. Thus the cellular automaton
acts  by applying the same unitary rotation separately to every
cell.
\item Fix a basis in $\Cx^d$ and
consider any unitary $U_0$ which is diagonal in the corresponding
product basis. Clearly, this guarantees that $U_0$ commutes with
its translates. That is, we can generate a QCA by an arbitrary
choice of {\it local phases} in some computational basis. A
prominent example is an Ising interaction
\begin{equation}\label{ising}
    H=(\idty-\sigma_3)\otimes(\idty-\sigma_3)
\end{equation}
turned on for a suitable finite time (e.g., $t=\pi/4$), which
generates the initial entangled state of a one-way quantum
computer \cite{onewaycomp} together with a cellwise Hadamard
rotation. \item Choose at every site a family of unitaries $V_p$,
$p=1,\ldots,d^2$ commuting up to a phase. Examples are the Pauli
matrices for $d=2$, or a discrete Weyl system. Then $U_0$ can be
any finite product of operators from this family, localized on
neighboring cells. The automata generated in this way form a
group.
\end{itemize}

\noindent For all QCAs constructed from commuting unitaries we
have the strange property that they show {\it no propagation}, in
the sense that the localization region does not increase when we
iterate the automaton. The reason is that $n$ steps are
implemented by a product of commuting unitaries, each of which
appears $n$ times. Hence it is exactly equivalent to take instead
a single step with the basic unitary operator $U_0$ replaced by
$U_0^n$.

Of course, most QCAs do have propagation. For example, we could
take a single step constructed from commuting unitaries, followed
by a site-wise rotation along a skew axis. Another rich class of
propagating QCAs is given in the next section.

\subsection{Quantization of Classical Reversible CAs}\label{sec:classical}
A typical feature of the local phase automata is that they leave
invariant the algebra of operators diagonal in the chosen basis.
This algebra $\DD$ is the quasi-local algebra of a classical CA
embedded into the quantum system, which has $d$ states per cell,
when $\AA_x$ is the algebra of $d\times d$-matrices. It is
therefore natural to look for cellular automata, which leave this
classical subalgebra $\DD$ invariant as a set, but not
elementwise. Clearly, such QCAs induce a classical CA on the
classical subsystem. In fact, {\it every} reversible classical CA
can be obtained in this way.

Intuitively, this is seen as follows: the classical CA can be run
with periodic boundary conditions, for simplicity, so we have only
a finite set $\loclat$ of cells. It then defines a permutation of
the $d^{|\loclat|}$ classical configurations, which we can
interpret immediately as a unitary permutation operator $U$. This
unitary operator is now used to implement the local evolution of
the QCA. All we need to verify is that for $A\in\AA_0$ the
evolution $U^*AU$ is indeed contained in some local algebra
$\AA(\Nei)$ for some regular neighborhood $\Nei$. Then the
wrapping Lemma asserts that the QCA is also well-defined on the
infinite lattice.

However, this argument must be more subtle than it looks: it is
well known, that the injectivity of a classical CA on the infinite
lattice implies the existence of an inverse CA \cite{Rich}, but
the inverse is hard to compute, because there is no a priori upper
bound on its neighborhood size. Superficially, the inverse
neighborhoods do not seem to enter the above argument. However,
the argument for the locality $U^*AU\in\AA(\Nei)$ requires more
than the locality of the classical rule and the unitarity of $U$.
Consider, for example, the rule
\begin{equation}\label{xorautomat}
 c^{t+1}_{x}=c^{t}_{x+1}+c^{t}_{x}+c^{t}_{x-1}\;,
\end{equation}
 where $c^{t}_{x}\in\{0,1\}$, and addition is mod$2$. With
periodic boundary conditions of length $L$ this is an invertible
transformation unless $L$ is divisible by 3. This proviso is not
of the form ``for sufficiently large $L$...'' , which means that
the classical automaton does not allow a local inversion, i.e.,
there is no inverse cellular automaton. By the wrapping Lemma, it
is clear that for this rule we cannot find a quantum version
either\cite{classicalW}. This shows that the {\it local}
invertibility of the classical CA must enter the argument.  We
therefore assume now that the classical CA is locally invertible,
and the lattice $\loclat$ is chosen sufficiently large, so that
the neighborhood schemes for both the classical automaton and its
inverse are regular for $\loclat$.

The classical configurations are functions
$\underline{a}:\loclat\to A$, where $A=\{1,\ldots,d\}$ denotes the
set of classical states for each cell, and at the same time labels
the computational basis of $\Cx^d$. We will write
$\underline{a}\in A^\loclat$, and denote by
$\underline{a}_x=\underline{a}(x)$ the value of the cell $x$ in
configuration $\underline{a}$.

Extending this to a product basis, each configuration
$\underline{a}\in A^\loclat$ determines a basis vector
$\ket{\underline{a}}$. Then the global unitary transition operator
is defined by $U\ket{\underline{a}}=\ket{F({\underline{a}})}$,
where $F$ denotes the global classical transition function. The
transition rule of the QCA is determined by computing all matrix
elements of the operator $\lrule(\ketbra{c_0}{e_0})$, i.e.,
\begin{eqnarray}\label{classmat}
  \bigl\langle \underline{a}\bigr|\lrule(\ketbra{c_0}{e_0}) \bigl|\underline{b}\bigr\rangle
  &=&\bigl\langle \underline{a}\bigr|U^*(\ketbra{c_0}{e_0}\otimes
                                          \idty^{\loclat\setminus\{0\}})U
        \bigl|\underline{b}\bigr\rangle
      \nonumber\\
  &=&\bigl\langle F(\underline{a})\bigr|(\ketbra{c_0}{e_0}\otimes
                         \idty^{\loclat\setminus\{0\}})
\bigl|F(\underline{b})\bigr\rangle
      \nonumber
\end{eqnarray}
This expression is $=1$, if
\begin{eqnarray}\label{transCCA}
    \mbox{if}\  F(\underline{a})_0&=&c_0,\quad
                F(\underline{b})_0=e_0,\ \nonumber\\
    \mbox{and}\ F(\underline{a})_x&=&F(\underline{b})_x
                 \quad\mbox{for}\  x\neq0
\end{eqnarray}
and $=0$ otherwise. We have to show that this is of the form
$X\otimes\idty^{\otimes\loclat\setminus\Nei}$.

\begin{lemma}\label{classlem}Let $F$ be a classical cellular automaton
with neighborhood scheme $\Nei_C$, which has an inverse automaton
with neighborhood scheme $\Nei_I$. Then $\grule$ as defined above
is a QCA with neighborhood scheme $\Nei=\Nei_C-\Nei_C-\Nei_I$.
\end{lemma}

Of course, this Lemma only gives an upper bound on the size of the
neighborhood scheme. Depending on the particular automaton, $\Nei$
may be much smaller.

\proof\/: We have to show that for all $e_0,c_0$, the operator
$\lrule(\ketbra{c_0}{e_0})$ is of the form
$X\otimes\idty^{\otimes\loclat\setminus\Nei}$ with
$X\in\AA(\Nei)$. This is equivalent to two conditions: on the one
hand the matrix elements $\bigl\langle
\underline{a}\bigr|\lrule(\ketbra{c_0}{e_0})
\bigl|\underline{b}\bigr\rangle$ must vanish, whenever $a_x\neq
b_x$ for some point $x\notin\Nei$, and, moreover, the value of the
matrix elements must be independent of $a_x$ for such $x$.

Suppose that the matrix element is non-zero, i.e.,
condition~(\ref{transCCA}) holds. Then for all $y$ such that
$y+\Nei_I\neq0$ the computation of the the values of $a_y$ and
$b_y$ from $F(\underline{a})$ and $F(\underline{b})$ will give the
same values of $a_y$ and $b_y$. In other words, the diagonality
condition holds for $y\notin(-\Nei_I)$.

The dependence of the matrix element on $a_x$ is also governed by
condition~(\ref{transCCA}): since the matrix elements can only be
$0$ or $1$, we have to show that the validity of the condition
does not depend on $a_x$ for $x\notin\Nei$, given the range of
equality of $a'$s and $b'$s established in the previous paragraph.
Indeed, once that diagonality is established, one can see that
most of the conditions~(\ref{transCCA}) become redundant. If $x$
is such that $(x+\Nei_C)\cap(-\Nei_I)=\emptyset$, then
$F(\underline{a})_x$ and $F(\underline{b})_x$ are computed by the
local rule of $F$ from identical data, so they must be equal. It
therefore suffices to consider the condition for those finitely
many $x$ for which a dependence remains possible, i.e.,
$x\in(-\Nei_C-\Nei_I)$. But then only those $a_y$ enter, which
contribute to $F(\underline{a})_x$ via the local rule (and
similarly for $\underline{b}$). This restricts $y$ to
$(\Nei_C-\Nei_C-\Nei_I)$, and this is what the Lemma claims as the
localization region. \qed

\subsection{Partitioning}
The easiest way to build a cellular automaton with readily
verified locality properties is based on the cellwise unitary
rotations, with the modification of changing the partitioning of
the system into cells \cite{Lloyd}. The typical construction would
thus be:
\begin{enumerate}
 \item possibly divide the given cells into suitable subcells, by
writing the one-cell Hilbert space $\Cx^d$ as a tensor product of
other spaces.
 \item partition the set of cells of the previous step into blocks in some periodic
 way: every cell now belongs to exactly one block, and any two blocks
 are connected by a lattice translation.
 \item apply the same unitary operator to each block algebra.
 \item possibly repeat this procedure with different block
       partitions
 \item possibly split and regroup once again to come back to the original pattern of
$d$-dimensional cells.
\end{enumerate}

Every step in this construction is well-defined for the global
system for the same reason that cell-wise rotations are valid QCA
operations. Moreover, the unitary operators doing each of the
steps are essentially arbitrary, and by just inverting the steps
we can immediately construct the inverse QCA. This is why
partitioned cellular automata are sometimes called {\it
structurally reversible}. Moreover, the partitioning idea allows
one to construct QCAs from {\it irreversible} local rules just as
easily as reversible ones \cite{Brennen,RiWe}.

\begin{figure}[htb]
{\epsfxsize=5cm \epsffile{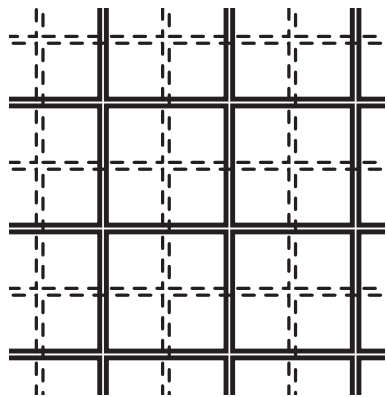}
 \caption{\label{margolus}{\it The Margolus partitioning scheme in $s=2$ dimensions.
 Operations are alternatingly applied to the solid and to the dashed
 partitioning into $2\times2$ squares. The square shape of the cells is irrelevant,
 as they only serve to label localized quantum systems.}}}
\end{figure}

Our main theorem (Theorem~\ref{mainthm} below) will tell us that
{\it every} QCA can be written in partitioned form. For nearest
neighbor automata it is sufficient to take two steps in which
cells are grouped in cubes of side 2, possibly with a choice of
unequal cell sizes in the intermediate step (see
Section~\ref{sec:margolus}). This is known as the Margolus
partitioning scheme (see Figure~\ref{margolus}).

\subsection{Circuits}

Of course, it is natural to think of a cellular automaton as a
physical device, which just happens to be infinitely extended (or
periodically closed). This suggests building QCAs from some basic
supply of circuit elements, whose properties will then ensure that
the overall operation makes sense. The mathematical details of the
description must then work out automatically, because ``Hardware
cannot lie''. Consider the following example

\begin{figure}[htb]
\epsfxsize=5cm \epsffile{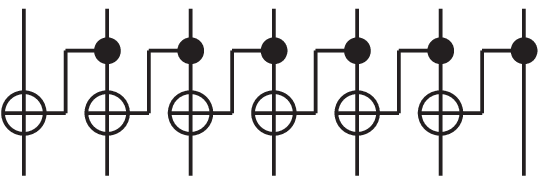}
 \caption{\label{circ1}{\it A proposed QCA circuit.}}
\end{figure}

Here the flow of information is from top to bottom. The symbol
stands for a CNOT gate, by which the bit on the line with circle
and cross (the ``target bit'') is flipped if and only if the value
of the bit on the line with the fat dot (the ``control bit'') is
``$1$''. This is a standard gate also for quantum computation. We
can readily compute the action of this device on classical
information: each output bit depends only on the input on the same
line and the line one step to the right. Presumably, this would
also be the description of the action on computational basis
states in the quantum case. So can we not just build this device,
and construct its proper mathematical description along with the
hardware?

The problem with this automaton becomes apparent already when we
try to compute its classical inverse: this requires at each switch
to know the value of a control bit, which is not yet determined.
In fact, the inverse does not exist, because the initial states
``all {\tt1}'' and ``all {\tt0}'' are both mapped to ``all
{\tt0}''. So this device is not reversible, even classically.
(Incidentally, this also holds for periodic boundary conditions,
so it is not a problem of infinite size).  What went wrong? The
problem is the {\it timing} of the gates. In fact, in the usual
gate model of quantum computation it is assumed that each gate is
executed at a certain time. Here the times overlap, and if we
insist on the gates being executed, say from the left to the
right, we either need infinitely many operation times per step, or
we run into time ordering problems at the boundary condition.

\begin{figure}[htb]
\epsfxsize=5cm \epsffile{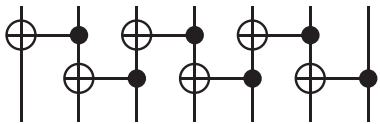}
 \caption{\label{circ2}{\it An operational QCA circuit
         taking two clock cycles}}
\end{figure}

One way to avoid this problem is to insist on some finite number
of clock cycles per QCA step, that each gate is executed in one of
these cycles, and that the gates running in the same cycle do not
access the same registers. An ``unscrambled version'' of the above
impossible QCA is drawn in Fig.~\ref{circ2}

Clearly, this is a partitioning QCA and, in fact, the description
we just gave of a circuit with timing constraints is nothing but
the definition of a partitioning QCA.

\subsection{Clifford automata}

In the implementations of quantum computation there is often a
separation between ``easily implemented'' operations and others,
which may be more costly. For example, ''linear'' transformations
on the quantum light field can be performed with mirrors, beam
splitters and phase plates, whereas squeezing or photon number
counting are more costly. A similar choice of a subgroup of
``easy'' operations for qubit quantum computation is the group of
transformations, which take tensor products of Pauli matrices into
tensor products of Pauli matrices, the so called {\it Clifford
group}\cite{cliff}. Indeed in some implementations these play a
special role, and can be executed in parallel \cite{onewaycomp}.
Clearly, it is important to understand this subgroup completely,
although it is also clear that such transformations alone will not
allow quantum computational speedup.

The natural mathematical setting for the investigation of Clifford
QCAs are discrete Weyl systems, with a finite Weyl system acting
at each site, and the tensor products of one-site Weyl operators
generating a Weyl system with infinitely many degrees of freedom.
If the local Weyl system has prime dimension $d$, one can give a
very explicit description of the group of Clifford QCAs, which
will be presented elsewhere \cite{OurCliff}. For illustration let
us just take qubit automata $d=2$ in one dimension.

What is needed to define such an automaton? Since the Pauli
matrices $\sigma_x$ and $\sigma_z$ generate the one site algebra,
we only need to specify the two operators $\lrule(\sigma_x)$ and
$\lrule(\sigma_y)$. The Clifford property means that these two
operators must be tensor products of Pauli matrices, so we have
\begin{eqnarray}\label{CQCAstring}
    \lrule(\sigma_x)&=&\sigma_{\xi_{-N}}\otimes\cdots\sigma_{\xi_{0}}
                               \otimes\cdots\sigma_{\xi_{N}}
                               \equiv\sigma(\xi)\\
    \lrule(\sigma_x)&=&\sigma_{\eta_{-N}}\otimes\cdots\sigma_{\eta_{0}}
                               \otimes\cdots\sigma_{\eta_{N}}
                               \equiv\sigma(\eta)\;,\nonumber
\end{eqnarray}
where each $\xi_i,\eta_i$ can take the values $0,x,y,z$, with
$\sigma_0=\idty$. Now the two strings
$\xi=(\xi_{-N},\ldots,\xi_{N})$ and
$\eta=(\eta_{-N},\ldots,\eta_{N})$ completely characterize the
automaton. But which strings are allowed? From the general theory
we immediately get the necessary and sufficient conditions:
$\sigma(\xi)$ must commute with all its translates, the same holds
for $\sigma(\eta)$. Moreover, $\sigma(\xi)$ commutes with
$\tau_i(\sigma(\eta))$ for $i\neq0$ and anti-commutes for $i=0$.
Since tensor products of Pauli matrices always either commute or
anti-commute, one can run a computer search for all examples with
low neighborhood size $N$. This turns up the surprising result
that (up to a common translation) the strings $\xi$ and $\eta$
must be {\it palindromes}, i.e., $\xi_{-k}=\xi_k$.

The general theory \cite{OurCliff} confirms this, and moreover
etablishes an isomorphism of the group of Clifford QCAs with the
group of $2\times2$-matrices,
\begin{equation}\label{CQCAmat}
    \left(\begin{array}{cc}
     \xi_+(z)&\eta_+(z)\\\xi_-(z)&\eta_-(z)
     \end{array}\right)\;,
\end{equation}
whose entries are polynomials over the two-element field
$\Field_2=\{0,1\}$ in one indeterminate $z$, such that the
determinate is the constant polynomial $1$. Here the coefficients
of $\xi_\pm$ are bit strings which together determine the string
$\xi_0,\xi_1,\ldots\xi_N$ used in Eq.~(\ref{CQCAstring}).

One can also find a simple set of generators: Apart from one-site
transformations and the shift only one QCA is needed:
\begin{eqnarray}\label{CQCAxzx}
    \lrule(\sigma_x)&=&\idty\otimes\sigma_{z}
                   \otimes\idty\\
    \lrule(\sigma_x)&=&\sigma_{z}\otimes\sigma_{x}\otimes\sigma_{z}\;,\nonumber
\end{eqnarray}
possibly, however, acting not between neighboring sites as written
here, but between sites at a fixed distance $L$, so that the chain
breaks up into $L$ non-interacting chains.

The prototype (\ref{CQCAxzx}) has a number of interesting
properties. For example, the iterates $T^t(\sigma(\zeta))$ of an
initial Pauli product $\sigma(\zeta)$ have a specific form: they
consist of a Pauli product moving to the left, and  another one
moving to the right, each at maximal speed, and the expanding
space between these patterns is filled by one of four
possibilities: all $\idty$, all $\sigma_y$ or an alternating
patterns of $\sigma_x\otimes\sigma_y$, or the same shifted by one
cell.

\section{Structure}\label{sec:struc}
In this section we will employ the commutativity property of
transition rules to get some information about the structure of
possible rules, aiming at the proof that all QCA transition rules
can be understood in a partitioning scheme.

Without loss of generality, we consider only nearest neighbor
rules on a cubic lattice. For a non-cubic lattice we can choose a
family of basic cells (a ``fundamental domain'') such that all
cells are generated from these basic ones by translation
symmetries. By considering the fundamental domain and its
translates as new cells, we effectively get a family of lattice
cells labelled by $\Ir^s$. If the neighborhood scheme involves
more than nearest neighbors, we can again enlarge cells. Of
course, these operations partly destroy the underlying lattice
symmetry, so that operations on regrouped cells may fail to have
the translation (or other) symmetry of the original lattice.
However, for the proof of structural invertibility a regrouped
lattice of ``supercells'' works just as well.

We begin by describing the geometry of the generalized Margolus
partitioning scheme. In the following subsection we state the main
theorem: this scheme indeed suffices for all QCAs. The proof
relies on the concept of ``support algebras'', and is described in
Subsection~\ref{sec:supp}. When the support algebras are abelian,
one can characterize the possible QCAs at the single cell level,
without partitioning (see Subsection~\ref{sec:abelianDD}). This
allows us to determine explicitly all nearest neighbor qubit
automata in one dimension (Section~\ref{sec:qubits}).

\subsection{Generalized Margolus Partitioning}
\label{sec:margolus}

\begin{figure}[htb]
\epsfxsize=5cm \epsffile{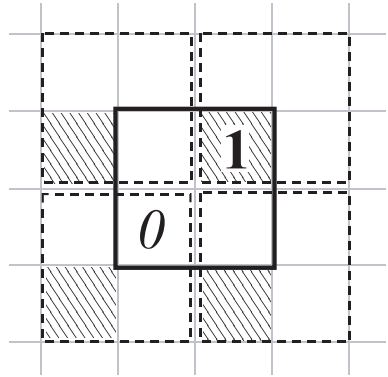}
 \caption{\label{margolusT}{\it The Margolus scheme in $s=2$ dimensions.
 The basic cube `` $\Box$'' is at the center, with the origin marked ``\,$0$''.
 The shaded squares are reached from the origin by quadrant vectors.
 Dashed outlines mark the copies $\Box+q$ shifted by quadrant vectors.}}
\end{figure}

Consider a cellular automaton with one-site algebra $\AA_0=\MM_d$,
lattice $\Ir^s$, and {\it nearest neighborhood}
scheme:
\begin{equation}\label{NN}
  \Nei=\{x\in\Ir^s|\  \forall_i|x_i|\leq1\}\;.
\end{equation}
We will use a cell grouping introduced by
Margolus \cite{MarTo}. In this scheme one ``supercell'' is the
unit cube
\begin{equation}\label{cube}
  \Box=\{x\in\Ir^s|\ \forall_i x_i=0,1\}\;,
\end{equation}
which consists of $2^s$ cells. The even translates $\Box+2x$ with
$x\in\Ir^s$ cover the whole lattice. We denote by $Q$ the set of
$2^s$ {\it quadrant vectors} $q\in\Ir^s$ for which each component
is $q_i\in\{-1,+1\}$ (see Fig~\ref{margolusT}). In particular the
vector into the positive quadrant (with all $q_i=+1$) will be
denoted by $\qone$. Note that the sum of two quadrant vectors is
an even lattice translation in $2\Ir^s$. Moreover, the $2^s$ cubes
$\Box+q$ for $q\in Q$ are disjoint and their union contains all
neighborhoods of cells in $\Box$. Just like the even translates of
$\Box$, the cubes $\Box+\qone+2x$ with $x\in\Ir^s$ form a
partition of the lattice (compare Fig.~\ref{margolus}).

A partitioned automaton based on the Margolus scheme would
alternatingly apply blockwise unitary transformations to the cubes
in the two partitions. But not every QCA can be written in this
way. A good counterexample is the shift in the quadrant direction
$\qone$. Here the entire quantum information in the $\Ir^s$ cells
$\Box$ will have to be moved to $\Box+\qone$ although these blocks
have only a single cell as overlap. It turns out, however, that a
slight generalization of the Margolus scheme suffices to represent
every QCA: all we have to do is to allow different cell sizes in
the intermediate step. For the rest of this subsection we will
explain the resulting scheme.

With each quadrant vector $q\in Q$ we associate an observable
algebra $\BB_q\subset\AA(\Box+q)$, which is isomorphic to the
algebra of $n(q)\times n(q)$-matrices for some integer $n(q)$.
Since $\BB_q$ is contained in a local algebra, it makes sense to
consider its (even) translates $\tau_x(\BB_q)\subset\AA(\Box+q+x)$
with $x\in2\Ir^s$. In particular, $\AA(\Box+\qone)$ contains all
the algebras $\tau_{\qone-q}(\BB_q)$. A crucial assumption of our
construction is that these subalgebras of $\AA(\Box+\qone)$
commute, and together span $\AA(\Box+q)$. This is possible if and
only if the equation
\begin{equation}\label{blockdims}
   \prod_{q\in Q}n(q)=\dtwos
\end{equation}
holds for the matrix dimensions. Note that each cell $\AA(\Box+x)$
has this dimension, whether or not $x$ is even or not.

The local rule of an automaton now defines (and is defined by) a
homomorphism
\begin{equation}\label{locMarg}
    \brule:\AA(\Box)\to\prod_{q\in Q}\BB_q\;.
\end{equation}
Since the dimensions of domain and range are the same, such a
homomorphism is necessarily an isomorphism, i.e., we can find
suitable bases in each cell and for each of the matrix algebras
$\BB_q$ such that $\brule(A)=UAU^*$ for a unitary operator
\begin{equation}\label{bruleU}
    U:\bigotimes_{x\in\Box}\Cx^d\longrightarrow\bigotimes_q
                       \Cx^{n(q)}\;.
\end{equation}
Such a unitary operator by itself does not fix a QCA, because if
we only take $\BB_q$ as an abstract matrix algebra, we still need
to specify how $\tau_{\qone-q}\BB_q$ is contained in
$\AA(\Box+\qone)$ or, equivalently,  to specify the isomorphism of
$\bigotimes_q\tau_{\qone-q}\BB_q$ with $\AA(\Box+\qone)$. This
will be affected by another unitary operator
\begin{equation}\label{bruleV}
    V:\bigotimes_q \Cx^{n(q)}\longrightarrow\bigotimes_{x\in\Box+\qone}\Cx^d\;.
\end{equation}

\begin{figure}[htb]
\epsfxsize=5cm \epsffile{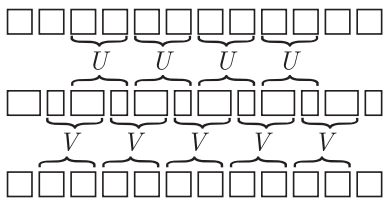}
 \caption{\label{margolus1}{\it Generalized Margolus Scheme in $s=1$ dimension.
 The algebras $\BB_{+1}$ and $\BB_{-1}$ are symbolized by the different size cells
 in the intermediate step.}}
\end{figure}

\begin{figure}[htb]
\epsfxsize=7cm \epsffile{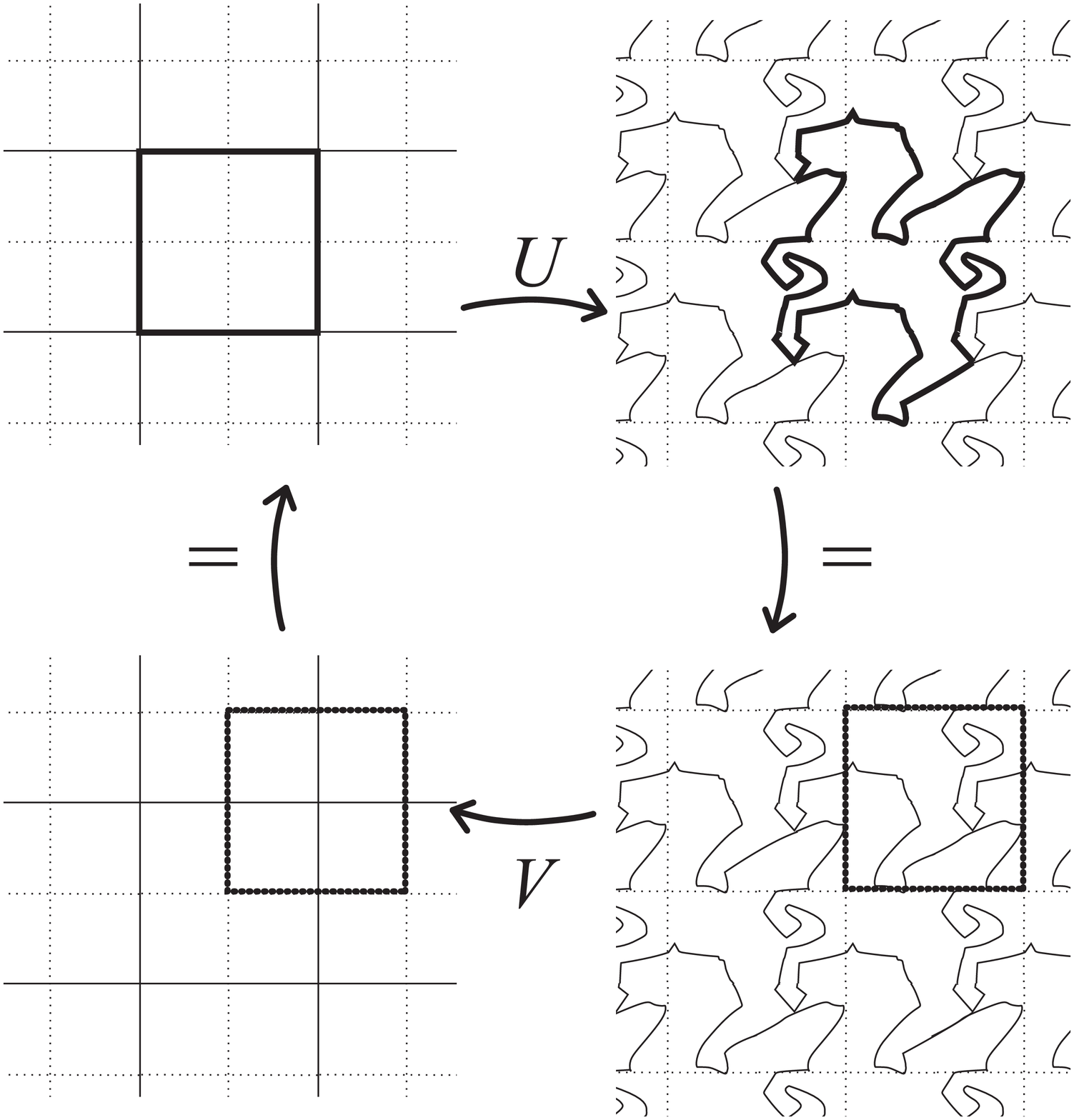}
 \caption{\label{pegasus}{\it Generalized Margolus Scheme in $s=2$ dimensions.
 The four subalgebras $\BB_q$ are symbolized by the shapes in Fig.~\ref{figchops}.
 The tesselation is due to M.C. Escher (1956). }}
\end{figure}

\begin{figure}[htb]
\epsfxsize=3cm \epsffile{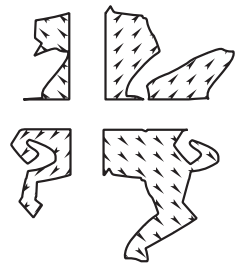}
 \caption{\label{figchops}{\it The four shapes representing the algebras
 $\BB_q$ in Fig.~\ref{pegasus}. The arrows indicate the appropriate quadrant vectors $q$.
}}
\end{figure}

Any pair of unitaries $(U,V)$ according to
Eqs.~(\ref{bruleU},\ref{bruleV}) specifies a transformation $T$ on
local algebras, which satisfies all requirements for a cellular
automaton, except translation invariance: $T$ only commutes with
even translations. The scheme in one and two lattice dimensions is
visualized in Figures~\ref{margolus1}-\ref{figchops}.

%
%\vskip40pt

If we want $\grule$, as constructed from unitaries $U$ and $V$, to
be a proper cellular automaton with full translation invariance,
there will be additional conditions on these unitaries.
Unfortunately, these conditions are not easily written down and
solved in the general case. We also note that $U$ and $V$ are not
uniquely determined by the automaton: we have the freedom to
choose a basis in every $\Cx^{n(q)}$. Changing this basis amounts
to a cellwise rotation included in $U$, which is immediately
undone by the $V$-step. The key feature of automata in a
partitioned scheme is {\it structural reversibility}, as described
in the following Lemma.

\begin{lemma}\label{invMarge}
Let $\grule$ be a homomorphism constructed from unitaries $(U,V)$
in the generalized Margolus scheme. Then $\grule$ is invertible,
and $\grule^{-1}$ is also a generalized Margolus automaton.
Moreover, if $\grule$ commutes with all (not just even)
translations, then both $\grule$ and $\grule^{-1}$ are nearest
neighbor QCAs.
\end{lemma}

\proof: The Margolus unitaries defining $\grule^{-1}$ are
$(V^*,U^*)$. To check the localization properties, it is helpful
not to think of these transformations in the apparently
time-asymmetric scheme of Fig.~\ref{pegasus}, but to take $\BB_q$
as an algebra localized in the intersection of the cubes $\Box$
and $(\Box+q)$, i.e., as localized at the cell $(\qone+q)/2$. Note
that $\grule^{-1}$ is a two-sided inverse, and hence uniquely
determined by $\grule$.

Therefore, if $\grule$ commutes with all translations, so does
$\grule^{-1}$. It remains to check that if $\grule$ is a Margolus
automaton commuting with translations, it is actually a nearest
neighbor automaton, i.e., a QCA with neighborhood scheme $\Nei$
from (\ref{NN}). Since, for every $x\in\Box$, we have
$0\in(\Box-x)$, we have
\begin{equation}
   \grule(\AA_0)
      \subset\grule(\AA(\Box-x))
      \subset\bigotimes_{q\in Q} \AA(\Box+q-x)\;.
\end{equation}
Since this is valid for {\it all} $x\in\Box$, we have
$\grule(\AA_0)\subset\AA(\widetilde\Nei)$, with
\begin{equation}\label{neimarg}
    \widetilde\Nei=\bigcap_{x\in\,\Box}\bigcup_{q\in Q}(\Box+q-x)=\Nei
\end{equation}
 \qed

\subsection{Main Theorem}

\vbox{
\begin{theorem}\label{mainthm}
Let $\grule$ be the global transition homomorphism of a nearest
neighbor quantum cellular automaton on the lattice $\Ir^s$ with
single-cell algebra $\AA_0=\MM_d$. Then $\grule$ can be
represented in the generalized Margolus partitioning scheme, i.e.,
$\grule$ restricts to an isomorphism
\begin{equation}
    \grule:\AA(\Box)\longrightarrow \bigotimes_{q\in Q}\BB_q \;,
\end{equation}
where for each quadrant vector $q\in Q$, the subalgebra
$\BB_q\subset\AA(\Box+q)$ is a full matrix algebra,
$\BB_q\cong\MM_{n(q)}$. These algebras and the matrix dimensions
$n(q)$, which satisfy Eq.~(\ref{blockdims}), are uniquely
determined by $\grule$.
\end{theorem}
}

Combining this with Lemma~\ref{invMarge}, we get

\begin{corr}\label{cor:invert}
The inverse of a nearest neighbor QCA exists, and is a nearest
neighbor QCA.
\end{corr}

Note that this result is in stark contrast to the classical
situation. In the classical case the inverse of an injective CA is
a CA, i.e., locally invertible, but it is a highly non-trivial
matter to determine the neighborhood scheme of the inverse, which
can be much larger than the neighborhood of the automaton itself.
This is not a contradiction with the observation that every
classical CA can be quantized (see Section~\ref{sec:classical}):
In order to construct a QCA from a classical CA, we needed the
neighborhood size of the inverse.

The proof of the Theorem will be given in
subsection~\ref{sec:proof}. The key idea is to  construct
$\BB_q\subset\AA(\Box+q)$ explicitly from the inclusion
$\grule(\AA(\Box))\subset\bigotimes_q\AA(\Box+q)$. This
construction, which will also be useful independently, will be
described in the next section.

\subsection{Support Algebras of Local Rules}\label{sec:supp}

By definition, the transition rule $\lrule$ maps one cell algebra
into a tensor product of neighboring ones. Therefore we need to
investigate just how one subalgebra can sit inside a tensor
product of others.

Consider a subalgebra $\AA\subset\BB_1\otimes\BB_2$ of tensor
product. For the moment let us forget about the multiplication
laws, and just consider these as vector spaces, with the tensor
product known from the (multi-)linear algebra of finite
dimensional vector spaces. Then we can expand each $a\in\AA$ into
a sum $a=\sum_\mu b_\mu^{(1)}\otimes b_\mu^{(2)}$. But we might
get by just using a small subset of operators $b_\mu^{(i)}$. The
smallest subspace of $\BB_1$ sufficient for these expansions will
be called the {\it support} of $\AA$ on the first factor, and will
be denoted by $\spp(\AA,\BB_1)$. For a more formal definition note
that each $a\in\AA$ can be expanded uniquely in the form
$a=\sum_\mu a_\mu\otimes e_\mu$, where $\{e_\mu\}$ is fixed a
basis of $\BB_2$. Then $\spp(\AA,\BB_1)$ is the linear span of all
$a_\mu$ in this expansion, and is clearly independent of the basis
$\{e_\mu\}$ chosen for $\BB_2$. Then it is clear that
$b_\mu^{(i)}\in \spp(\AA,\BB_i)$ indeed suffice to expand every
$a\in\AA$, i.e.,
\begin{equation}\label{a2supp}
  \AA\subset \spp(\AA,\BB_1)\otimes \spp(\AA,\BB_2)
      \subset\BB_1\otimes\BB_2\;.
\end{equation}
The analogous relation for $\AA$ contained in a tensor product of
more factors is seen in the same way.

Now we remember the algebraic structure: $\spp(\AA,\BB_i)$ is a
linear subspace of a C*-algebra, so we can define the {\it support
algebra} $\Spp(\AA,\BB_i)$ of $\AA\subset \bigotimes_i\BB_i$ on
one factor $\BB_i$ as the subalgebra of $\BB_i$ generated by the
elements of $\spp(\AA,\BB_i)$ \cite{Zanardi}. Then we also have
\begin{equation}\label{a2Supp}
  \AA\subset \Spp(\AA,\BB_1)\otimes \Spp(\AA,\BB_2)
      \subset\BB_1\otimes\BB_2\;.
\end{equation}
 Note that since any observable algebra $\AA$ is closed under adjoints, so is
$\Spp(\AA,\BB_1)$. The crucial fact we need about such inclusions
is the following:

\begin{lemma} \label{sppcomm}
Let $\AA_1\subset\BB_1\otimes\BB_2$ and
$\AA_2\subset\BB_2\otimes\BB_3$ be subalgebras such that
$\AA_1\otimes\idty_3$ and $\idty_1\otimes\AA_2$ commute in
$\BB_1\otimes\BB_2\otimes\BB_3$. Then $\Spp(\AA_1,\BB_2)$ and
$\Spp(\AA_2,\BB_2)$ commute in $\BB_2$.
\end{lemma}

\proof:\quad Pick bases $\{e_\mu\}\subset\BB_1$ and
$\{e'_\nu\}\subset\BB_2$, and let $a\in\AA_1$ and $a'\in\AA_2$.
Then we may expand uniquely: $a=\sum_\mu e_\mu\otimes a_\mu$ and
$a'=\sum_\nu a'_\nu\otimes e'_\nu$. Then by assumption
\begin{displaymath}
 0=[a\otimes\idty_3,\idty_1\otimes a']
  =\sum_{\mu\nu}e_\mu\otimes [a_\mu,a'_\nu]\otimes e'_\nu \;.
\end{displaymath}
Now since the elements $e_\mu\otimes e'_\nu$ are a basis of
$\BB_1\otimes\BB_3$, this expansion is unique, so we must have
$[a_\mu,a'_\nu]=0$ for all $\mu,\nu$. Clearly, this property also
transfers to the algebras generated by the $a_\mu$ and $a'_\nu$,
i.e., to the support algebras noted in the Lemma. \qed

\subsection{Proof of the Main Theorem}\label{sec:proof}

We apply the construction of support algebras to the inclusion
\begin{equation}\label{inclusion}
  \grule(\AA(\Box))\subset\bigotimes_q\AA(\Box+q)
\end{equation}
and define
\begin{equation}\label{defBBq}
    \BB_q=\Spp(\grule(\AA(\Box)),\AA(\Box+q))\;.
\end{equation}

As a finite dimensional C*-algebra, each $\BB_q$ is isomorphic to
$\BB_q=\bigoplus_\mu\MM_{n(q,\mu)}$ (See
Proposition~\ref{Csform}). Now $\grule(\AA(\Box))$ is
homomorphically embedded into $\bigotimes_q\BB_q$, with
$\grule(\idty)=\idty$. Hence by Proposition~\ref{Cshom} we know
that for any choice of summands $\mu_q$ we must have that
$\dtwos$, the matrix dimension of $\AA(\Box)$, divides
$\prod_qn(q,\mu_q)$. This gives a lower bound on the block sizes.

In order to get an upper bound, consider the support algebras
contained in some shifted cube, such as $\Box+\qone$. These are
\begin{eqnarray}
 \Spp\Bigl(\grule(\AA(\Box+\qone-q))&,&\AA(\Box+\qone)\Bigr)
  \nonumber\\
  &=&\tau_{\qone-q}\Spp\Bigl(\grule(\AA(\Box)),\AA(\Box+q)\Bigr)
  \nonumber\\
  &=&\tau_{\qone-q}(\BB_q)\;
\end{eqnarray}
Since the $\AA(\Box+\qone-q)$ commute, so do their images under
$\grule$ and, by Lemma~\ref{sppcomm}, so do the algebras
$\tau_{\qone-q}(\BB_q)\subset\AA(\Box+\qone)$. However, we do not
know a priori that these algebras are contained in
$\AA(\Box+\qone)$ like a tensor product: When
$z_{q,\mu_q}\in\BB_q$ is a central projection onto one of the
matrix blocks of $\BB_q$, it is clear that the product
$\prod_qz_{q,\mu_q}$ is a central element of the algebra generated
by the $\BB_q$, but it might be zero. On the other hand, there
must be {\it some} choice of blocks $\mu_q$, for which this is
non-zero, and for this combination $\prod_q\MM_{n(q,\mu_q)}$,
which is isomorphic to $\MM_n$ with $n=\prod_qn(q,\mu_q)$, is a
direct summand of $\prod_q\BB_q\subset\AA(\Box+\qone)$. Hence we
get the inequality
\begin{equation}\label{dimineq}
   \dtwos\geq\prod_qn(q,\mu_q)\;.
\end{equation}

On the other hand, by the first step, the left hand side divides
the right hand side of this inequality, so the we must have
equality. This also implies that only one summand can be present
in $\BB_q$, so we get $\BB_q\cong\MM_{n(q)}$ with
$n(q)=n(q,\mu_q)$. That $\grule$ is an isomorphism from
$\AA(\Box)$ onto $\bigotimes_q\BB_q$ follows by a direct dimension
count, since a *-homomorphism between full matrix algebras of
equal dimension can only be zero or an isomorphism. \qed

\subsection{QCAs with abelian neighborhood}\label{sec:abelianDD}
In a sense, Theorem~\ref{mainthm} gives a complete constructive
procedure for QCAs in terms of the two unitary operators $U$, $V$
with a constraint. Unfortunately, however, it does not seem to be
easy to give a general solution of  the constraint equations
expressing the translation invariance (rather than the invariance
by even translations). Therefore it is suggestive to repeat the
analysis of support algebras also on the single cell level.
Setting
\begin{equation}\label{Dlocal}
  \DD_x=\Spp\bigl(\lrule(\AA_0), \AA_x\bigr)
\end{equation}
we have
\begin{equation}\label{Dlocalsub}
  \lrule(\AA_0) \subset\bigotimes_{x\in\Nei}\DD_x
\end{equation}
It is clear that since $\lrule(\AA_0)$ is non-abelian, at least
one of the algebras $\DD_x$ must also be non-abelian. The simplest
case in this regard will be when all $\DD_x$ are abelian and
commute with each other, except one, say $\DD_0$, which then has
to isomorphic to the full cell algebra $\AA_0$ by
Prop.\ref{Cshom}. Since all $\DD_x$ commute, we can jointly
diagonalize them and this fixes a canonical basis for every cell.
When $\ket\mu$ denotes the basis vectors, we can write the local
transition rule as
\begin{equation}\label{conditionalrule}
  \lrule(A)
  =\sum_{\mu_\Nei} U(\mu_\Nei)^*AU(\mu_\Nei)\otimes
    \bigotimes_{0\neq x\in\Nei}\ketbra{\mu_x}{\mu_x}\;,
\end{equation}
 where the sum runs over all tuples $\mu_\Nei$ of basis labels
$\mu_x$ for $x\in\Nei, x\neq0$ and, for each such tuple, $U(\mu_\Nei)$ is
a unitary operator. Thus $\lrule$ describes a {\it conditional unitary
operation} on cell $0$, where the conditioning is in some fixed
``computational basis''.

We now need to analyze the constraints on these unitaries needed to make
this homomorphism $\lrule$ a local transition rule. As a first step we
look at the simplest case:

\begin{lemma} Let $U_\mu,V_\nu\in\MM_d$ be unitary operators ($\mu,\nu=1,\ldots,d$)
such that, for all $A,B\in\MM_d$,
\begin{equation}\label{condscommute}
   \sum_{\mu\nu}\Bigl[U_\mu^*AU_\mu\otimes\ketbra\mu\mu,\;
                      \ketbra\nu\nu\otimes V_\nu^*BV_\nu\Bigr]=0
\end{equation}
Then there are unitary $U,V\in\MM_d$ such that, for all $\mu$,
$U^*U_\mu$ and $V^*V_\mu$ are diagonal.
\end{lemma}

\proof :\quad Let us abbreviate $A_\mu=U_\mu^*AU_\mu$,
$B_\nu=V_\nu^*BV_\nu$, and take the matrix element of equation
(\ref{condscommute}) in the product basis vectors
$\bra{\alpha\beta}\cdots\ket{\gamma\delta}$. This gives
\begin{equation}
  \bra\alpha A_\beta\ket\gamma\; \bra\beta B_\gamma\ket\delta
  =\bra\alpha A_\delta\ket\gamma\; \bra\beta B_\alpha\ket\delta
  \nonumber
\end{equation}
 Now set $B=V_\alpha\ketbra{\beta'}{\delta'}V_\alpha^*$, with
 $\beta'\neq\beta$. Then $B_\alpha=\ketbra{\beta'}{\delta'}$, and
 $\bra\beta B_\alpha\ket\delta=0$, and the right hand side vanishes.
Hence, for every $A$ and every $\delta'$
\begin{equation}
  \bra\alpha A_\beta\ket\gamma\;
  \bra\beta V_\gamma^* V_\alpha\ket{\beta'}
  \bra{\delta'}V_\alpha^*V_\gamma \ket\delta
  =0.
  \nonumber
\end{equation}
Now the first factor can be made non-zero by an appropriate choice
of $A$, and the third factor can be made non-zero by choosing
$\delta'$, because the unitary operator $V_\alpha^*V_\gamma$
cannot annihilate $\ket\delta$. It follows that
 $\bra\beta V_\gamma^* V_\alpha\ket{\beta'}=0$ vanishes for all
indices, or $V_\gamma^* V_\alpha$ is diagonal for all
$\alpha,\gamma$. Hence the Lemma follows with $V=V_1$, and the
statement for $U$ follows by symmetry.\qed

Let us apply this Lemma to the one-dimensional nearest neighbor
case. Then a local rule of the form (\ref{conditionalrule}) can be
written as
\begin{equation}\label{conditionalrule1D}
  \lrule(A)
  =\sum_{\mu\nu} \ketbra\mu\mu\otimes U_{\mu\nu}^*AU_{\mu\nu}
                    \otimes\ketbra{\nu}{\nu};.
\end{equation}
 The remaining commutation condition is, for arbitrary one-site
 observables $A,B$,
\begin{eqnarray}\label{commute1D}
  0&=&[\lrule(A)\otimes\idty, \idty\otimes\lrule(B)]
                                \nonumber\\
  &=&\sum_{\mu\nu\mu'\nu'} \ketbra\mu\mu\otimes
    \Bigr[U_{\mu\nu}^*AU_{\mu\nu}\otimes\ketbra{\nu}{\nu},
                                    \nonumber\\ &&\qquad
      \ketbra{\mu'}{\mu'}\otimes U_{\mu'\nu'}^*BU_{\mu'\nu'}
    \Bigr]\otimes\ketbra{\nu'}{\nu'}
\end{eqnarray}
Hence we can apply the Lemma to the commutator separately for
every pair $\mu,\nu'$, and we find that up to a common unitary all
$U_{\mu\nu}$ have to commute. Again, up to a cell-wise unitary
rotation, we can choose the common eigenbasis of the $U_{\mu\nu}$
as the same basis in which the conditions are written, i.e.,
\begin{equation}\label{phase1D}
  U_{\mu\nu}=\sum_\kappa u(\mu\kappa\nu) \ketbra\kappa\kappa \;,
\end{equation}
 with some phase function $u$ depending on three neighboring basis
labels. However, this phase function $u$ is not arbitrary:
unitaries of the form (\ref{phase1D}) in this way may still fail
to satisfy (\ref{commute1D}). If we insert $A=\ketbra ab$ and
$B=\ketbra{a'}{b'}$ we get the functional equation
\begin{equation}\label{uuuu}
   \frac{u(\mu b a')}{u(\mu a a')}\cdot \frac{u(bb'\nu')}{u(ba'\nu')}
  =\frac{u(\mu b b')}{u(\mu a b')}\cdot \frac{u(ab'\nu')}{u(aa'\nu')}\;.
\end{equation}
Since we want to classify solutions up to a cell-wise rotation, we
can take one of the unitaries $U_{\mu\nu}$ to be the identity, say
$U_{11}=\idty$, or $u(1x1)=1$. Moreover, an overall phase of
$U_{\mu\nu}$ is irrelevant, and we can choose this so
$u(\mu1\nu)=1$. Then in (\ref{uuuu}) we take $a=b'=\nu'=1$, which
gives
\begin{equation}\label{uuuusolve}
  u(\mu ba')=u(\mu b1)u(ba'1)
\end{equation}
 Thus $u$ is already determined by the two-variable function
$(a,b)\mapsto u(a,b,1)$. It is easy to check that any choice of
this function yields a solution of (\ref{uuuu}) via
(\ref{uuuusolve}).

We can summarize the result as follows:

\begin{proposition}\label{prop:abelianDD}
 Let $\lrule$ be the local transition rule of
a QCA such that $\DD_1$ and $\DD_{-1}$ are both abelian. Then,
with respect to a basis in which these algebras are diagonal,
there is a phase gate on $\Cx^d\otimes \Cx^d$:
\begin{equation}\label{phasegate}
  U=\sum_{ab}u(a,b)\ketbra{ab}{ab} \;,
\end{equation}
 normalized such that $u(1,b)=u(b,1)=1$, and a one-site unitary
 $V$ such that
\begin{eqnarray}\label{phasealfa}
 \lrule(A)&=&X^*(\idty\otimes V^*AV\otimes\idty)X \qquad \mbox{with}\nonumber\\
  X&=& (U\otimes\idty_3)(\idty_1\otimes U)\;. \nonumber
\end{eqnarray}
\end{proposition}

\subsection{Unilateral Automata}

Another case of automata in one dimension, which can be characterized
completely, are automata with neighborhood scheme $\Nei=\{0,1\}$ (or,
symmetrically, $\Nei=\{-1,0\}$). The analysis is almost identical to that
of Theorem~\ref{mainthm}, and gives full matrix algebras
$\DD_i=\Spp(\lrule(\AA_0),\AA_i)=\MM_{n_i}$ and such that $n_0n_1=d$. As
in the case of the Theorem, the local rules combines an arbitrary unitary
$U:\Cx^d\to\Cx^{n_0}\otimes\Cx^{n_1}$ with a unitary
$V:\Cx^{n_1}\otimes\Cx^{n_0}\to\Cx^d$.  We only mention these to state
the following classification of the simplest case:

\subsection{Nearest neighbor qubit automata in one
dimension}\label{sec:qubits}

Consider the right and left support algebras
$\DD_{-1},\DD_{0},\DD_{+1}$ as in (\ref{Dlocal}). These are
subalgebras of the $2\times2$-matrices, which leaves three
possibilities: each of these algebras can either be trivial
($\DD_i=\Cx\idty$), an abelian two-state algebra (isomorphic to
the diagonal matrices, or the full algebra $\MM_2$.

Suppose that at least one of the algebras  $\DD_{\pm1}$, say
$\DD_{-1}$, is trivial. Then we have a unilateral automaton. Since
$n_0n_1=2$ we must have either $n_0=2, n_1=1$, a cell-wise unitary
rotation or $n_0=1, n_1=2$ a right shift, possibly combined with a
cell-wise rotation.

Suppose that none of the algebras $\DD_{\pm1}$ is trivial. Then
because $\DD_{-1}$ commutes with $\DD_{+1}$ neither algebra can be
the full matrix algebra, since that would force the other to be
trivial. It follows that both are abelian, and commute. Hence
after a basis change (by another cell-wise rotation) we can take
$\DD_{-1}=\DD_{+1}$ as the algebra of diagonal
$2\times2$-matrices. This brings us into the situation of
Section~\ref{sec:abelianDD}, and we find a phase rotation.
Choosing the normalization in Proposition~\ref{prop:abelianDD}, we
have only one free parameter left, i.e., we have an automaton
built from commuting unitaries $U_x$ (cf.
Section~\ref{sec:basicon}), which are {\it phase gates}
\begin{equation}\label{phigate}
  U_x=\left(\begin{matrix}1&0&0&0\\0&1&0&0\\0&0&1&0\\
             0&0&0&e^{i\phi}
       \end{matrix}\right)\;.
\end{equation}

\begin{figure}[htb]
\epsfxsize=6cm \epsffile{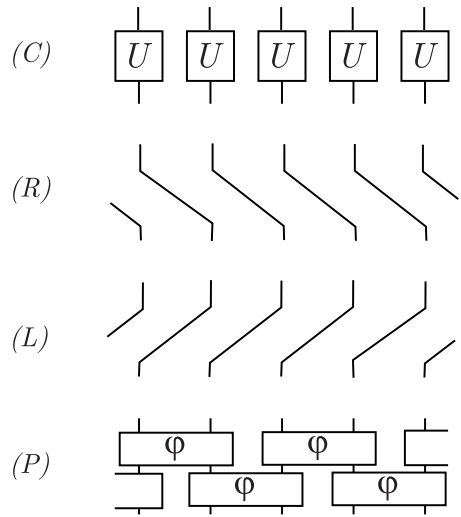}
 \caption{\label{qbit}{\it All nearest neighbor qubit automata arise by combining
  cell-wise unitary rotations $(C)$ with right shifts  $(R)$, left shifts $(L)$,
  or phase gates $(P)$.}}
\end{figure}

The classification is hence complete. It is represented in
Fig.\ref{qbit}.

It is interesting to compare this with the classical case, which
can be stated very similarly: then the only cell-wise operations
are identity and global flip. Of course, the possibility of phase
gates does not arise, leaving 6 classical possibilities.

Beyond qubits, a good classification exists for Clifford automata
in prime dimension \cite{OurCliff}. However, for general
three-level systems the classification is likely to be
complicated, since there is already a host of reversible classical
nearest neighbor automata.

\section{Other Approaches}\label{sec:others}
In this section we take a look at the various proposals for
defining QCAs, and show how they relate to the approach taken in
this paper.

\subsection{Feynman, and Transition Quasi-Probabilities}\label{sec:Feynman}
The idea of a quantum cellular automaton is clearly present in
Feynmans's famous 1981 lecture \cite{Feynman}. Although he
suggests not only that a theory might and should be developed, and
that it might even be taken seriously as fundamental physical
theory, he does not actually develop a notion of QCAs in this
article. The context in which he does write down a transition rule
for a QCA, is where he discusses the possibility of simulating a
QCA with a probabilistic cellular automaton. He emphasizes that
this would involve something like negative transition
probabilities and closes this section saying ``... I wanted to
explain that if I try my best to make the equations look as near
as possible to what would be imitable by a classical probabilistic
computer, I get into trouble''.

Feynman's idea for making a QCA look as classical as possible is
to replace ``transition probabilities'' by Wigner function-like
``transition quasi-probabilities''. This approach is an
interesting contribution to the definition of {\it irreversible}
QCAs, which is still plagued with problems. However, as Feynman is
clearly aware, it fails, and it is instructive to analyze this
failure in the case of reversible automata.

The transition function of a classical probabilistic cellular
automaton (for a set $S$ of single-cell states) is a set of
probabilities $M(s'|s_{\Nei})$ specifying the probability of
finding a state $s'\in S$, when the configuration of the neighbors
at the previous time step is $s_\Nei$. Using the ``simultaneous
and independent update rule'' the probability for finding a
configuration $s_\Lambda$ in a finite region $\Lambda\subset\Ir^s$
becomes
\begin{equation}\label{PCA}
  M(s'_\Lambda|s)=\prod_{x\in\Lambda}M(s'_x|s_{\Nei+x})\;.
\end{equation}
Note that this is readily read as a statement in the Heisenberg
picture: the probability for $s'_\Lambda$ is expressed as the
expectation of a random variable in the previous time step, namely
the right hand side of (\ref{PCA}), considered as a function of
the variables $s_x$. In the quantum case, $M(s'|s_\Nei)$ would
then become an observable in $\AA_\Nei$, depending in a linear
(and completely positive) way on an observable $s'\in\AA_0$. This
is precisely a description of the local transition rule $\lrule$.
Note that the commutation condition for local rules
(Lemma~\ref{lem:localrule}) implies that the product is
well-defined, independently of the ordering of the factors. In a
more general quantum context, such as irreversible QCA evolutions
for which the global rule will not respect the product,  we cannot
be sure of this property: additional information about the
ordering of factors in (\ref{PCA}) would have to be supplied, but
even an ordering fixed by some convention would not prevent the
evolution from sometimes taking hermitian elements to
non-hermitian elements.

This latter problem: the ordering of factors and the hermiticity
is neatly solved by Feynman's approach of quasi-probabilities. He
expands all qubit operators in a special basis of four hermitian
operators $F(\xi)\in\MM_2$, $\xi\in\{{++},{+-},{-+},{--}\}$:
\begin{equation}\label{Feyman_Wigner}
  F_{uv}=\left(\begin{matrix}
             (1+u)/2  &v(1-iu)/4\\
             v(1+iu)/4& (1-u)/2
          \end{matrix}\right) \;, (u,v=\pm1)\;.
\end{equation}
The expectations $f_\rho(\xi)=\tr(\rho F(\xi))$ of these operators
are the analogs of the Wigner function (see \cite{Wootters,Paz}
for later elaborations on Wigner functions in finite dimension).
Of course, by taking tensor products of these operators we get
Wigner functions for multi-qubit systems. Now we can expand the
transition rule for a single cell in Wigner operators:
\begin{equation}\label{alfa0Wig}
  \lrule(F(\eta_0))
    =\sum_{\xi_\Nei}M(\eta_0|\xi_\Nei)
      \bigotimes_{x\in\Nei} F(\xi_x) \;,
\end{equation}
 with the transition quasi-probabilities $M(\eta_0|\xi_\Nei)$.
These are usually not positive, but this is a minor inconvenience
as long as the physical transition operator $\grule$ is positive.
Equation~(\ref{alfa0Wig}) is just an expansion of the local rule
in a basis of Wigner operators, so a local rule is completely
equivalent to a set of transition quasi-probabilities. The
difference between the quasi-probability approach and ours is how
the global rule is constructed. Let us consider, for simplicity,
the image of a two-site observable $F(\eta_1)\otimes F(\eta_2)$
under a one-dimensional nearest neighbor automaton. According to
(\ref{PCA}), transition probabilities must be combined as
\begin{equation}\label{alfa1Wig}
\begin{array}{l}\displaystyle
  \grule_{\rm quasi}(F(\eta_1)\otimes F(\eta_2))\\ \displaystyle
   {}\quad=\sum_{\xi_0,\xi_1,\xi_2,\xi_3}M(\eta_1|\xi_0,\xi_1,\xi_2)
             M(\eta_2|\xi_1,\xi_2,\xi_3)
   \rule[-25pt]{0pt}{45pt}\\
   \displaystyle{}\ \qquad\qquad\times
    F(\xi_0)\otimes F(\xi_1)\otimes F(\xi_2)\otimes F(\xi_3)\;.
\end{array}
\end{equation}
 We can extend this to arbitrarily large configurations to get the
time evolution of an automaton in the Heisenberg picture. Since
the transition quasi-probabilities are real, this evolution will
automatically preserve hermiticity, and since $\sum_\xi
F(\xi)=\idty$ all normalization and locality properties will
automatically come out correctly. Of course, there will be no
operator ordering problem, since we multiply at the level of
functions, and all this will work for probabilistic irreversible
and reversible transition rules alike.

On the other hand, under the homomorphism $\grule$ the tensor
product $F(\eta_1)\otimes F(\eta_2)$ is evolved to
\begin{equation}\label{alfa2Wig}
\begin{array}{l}\displaystyle
  \grule(F(\eta_1)\otimes F(\eta_2))\\ \displaystyle
    {}\ = \sum_{{\xi_0,\xi_1,\xi_2,\xi_3}\atop{\xi'_0,\xi'_1,\xi'_2,\xi'_3}}
             M(\eta_1|\xi_0,\xi_1,\xi_2)
             M(\eta_2|\xi'_1,\xi'_2,\xi'_3)
    \rule[-25pt]{0pt}{45pt}\\\displaystyle
           {}\quad \times
    F(\xi_0)\otimes F(\xi_1)F(\xi'_1)
       \otimes F(\xi_2)F(\xi'_2)\otimes F(\xi'_3)
\end{array}
\end{equation}
Then (\ref{alfa1Wig})=(\ref{alfa2Wig}), iff
$F(\xi)F(\eta)=\delta_{\xi\eta}F(\xi)$, which is obviously true
for the minimal projections in a classical observable algebra, but
obviously false for the Wigner operators. So the automata studied
in this paper are not (or not in general) representable as
quasi-probabilistic CAs.

Feynman uses the approach based on (\ref{alfa1Wig}) only as a
caricature of a quantum process. He points out that negative
transition quasi-probabilities exclude a simulation of the global
time step (\ref{PCA}) by successive independent random trials. For
him this indicates the increased complexity of quantum
computation.

Whether or not this approach  works as a definition of (possibly
also irreversible) QCAs hinges on the question of positivity.
Non-positive transition probabilities would not be serious if they
lead to the construction of a legitimate (i.e., completely
positive) global transition rule. Unfortunately, however, there is
no indication that quantum positivity improves in the passage from
local to global rule. This can be seen already for a simple phase
gate array, i.e., (P) in Fig.~\ref{qbit}. For generic phase angle
$\varphi$ neither the local transition nor the global transition
e.g., the expression (\ref{alfa2Wig}), have positive transition
quasi-probabilities, although, of course, they correspond to
completely positive operations. On the other hand, the two-site
transition rule (\ref{alfa1Wig}) takes some positive operators
into non-positive ones.

An interesting special case occurs for phase angle $\phi=\pi$. In
that case, the local rule does give rise to positive, and even
{\it deterministic} transition quasi-probabilities.  They belong
to a classical deterministic CA, which like the phase gate QCA is
its own inverse. But if it is applied as in (\ref{alfa1Wig}), it
violates positivity.

\subsection{Watrous et al.}
One of the first serious attempts at the definition of QCAs was by
Watrous \cite{Watrous}. It is based on another ``quantization'' of
the transition probability formula (\ref{PCA}), inspired by
Feynman's notion that in quantum theory one must replace
probabilities by amplitudes. Thus one tries a product formula like
(\ref{PCA}) for the transition amplitudes, presumably defining in
this way the unitary transition operator for the whole process:
\begin{equation}\label{watrousCA}
 U(a|b)=\prod_x  u(a_x|b_{\Nei+x})\;,
\end{equation}
 where $a,b$ are classical configurations, labelling the basis
states of the QCA, and $b_{\Nei+x}$ is the configuration $b$
restricted to the neighborhood of $x$. The function $u$ plays the
role of the local transition rule. Basically the same definition
is also used in van Dam \cite{vanDam}, where it is phrased as an
assignment of a a product vector to every basis state in the
computational basis. Further work in this approach is to be found
in \cite{santa}, and in the textbook \cite{Gruska}, or a recent
introduction \cite{Aoun}. For the sake of discussion let us call
an automaton defined by (\ref{watrousCA}) a WQCA.  There are
several problems with this formula:

\begin{enumerate}
\item
 The infinite product may not be defined. This may be resolved by
either introducing a ``quiescent state'' which is invariant under
the evolution, and considering only superpositions of
configurations which are quiescent outside a finite region
\cite{Watrous,santa}, or to look at periodic boundary conditions
only \cite{vanDam}. Either approach works, but none of these
workarounds is necessary in our definition.

\item $U$ from (\ref{watrousCA}) has no reason to be unitary, and
most of the time it isn't. In other words, there is no
straightforward way of characterizing those local transition
amplitudes $u$ for which the formula does indeed define an
isometric (or stronger: a unitary) operator, in which case the
rule is called ``well-formed'' (or unitary). Whereas positivity
and normalization of the local rule $M$ on the right hand side of
(\ref{PCA}) guarantee the corresponding properties for the global
evolution, no equally simple criterion exists for well-formedness
(but see \cite{santa} for algorithms in the one-dimensional case).

\item Many unitary operators are not of the form
(\ref{watrousCA}). To begin with, the definition depends on the
choice of a preferred basis. In general, the product of two WQCA
unitaries need not be a WQCA (with larger neighborhood), and the
inverse of a WQCA may fail to be a WQCA. This
makes it hard to build a general theory on this definition.
\newline
 To get an example of these phenomena note that a necessary
condition for a unitary of Watrous form is that the computational
basis states are mapped to product states. Moreover, if we follow
the unitary operator by a site-wise unitary rotation, or precede
it by a product of phase gates (i.e., introducing additional phase
factors independent of the output labels $a_x$), we stay in this
class. But now consider a sitewise Hadamard rotation, followed by
a product of phase gates. It is easy to verify that such a map
does not take the computational basis states to product states (or
Briegel's one-way computers would not work.) So this is not a
WQCA, although it is the product of two WQCAs, and its inverse is
also a WQCA .

\item Even in the cases where the formula does work, such as,
e.g., for the multiplication by local phases, the size of the
neighborhood in (\ref{watrousCA}) is not the neighborhood scheme
describing the propagation of observable effects (the $\Nei$ of
our definition), but rather the $\widetilde\Nei$ from
Section~\ref{sec:communitary}. In fact, it is not even clear
whether a WQCA is necessarily local in our sense: We did not
manage to exclude the possibility that a local measurement after
one time step might allow inferences about arbitrarily distant
modifications of the state.

\item in the original paper \cite{Watrous}, Watrous also looks at
partitioned automata. Since we prove that all QCAs can be obtained
by a partitioning scheme, it might seem that WQCA$\supset$QCA, in
seeming contradiction to the Example in 3 above. However, Watrous
uses only a special case of partitioning, namely a unitary
followed by a permutation of subcells. Only if we extend the class
of WQCAs and include all their products, we get all QCAs.

\end{enumerate}
To summarize: Judged by the criteria in the introduction, the
notion of Watrous is not a satisfactory formalization of the idea
of quantum cellular automata. What is lacking is the easy passage
from local to global rules, and from global axiomatic to local
constructive description. For all problems involving the iteration
of automata, the fact that the class is not closed under
composition and inverse, puts a premature end to studies based on
WQCAs.

\subsection{Richter and Werner}

As mentioned earlier, the observable-based approach underlying our
definition was first used in \cite{RiWe} by one of us, but with a
focus on the irreversible case. In that paper partitioning
(dissipative cell evolution combined with permutation of subcells)
was used to allow a free construction satisfying a global locality
condition.

It is not so clear what should replace the axiomatic definition in
the irreversible case. A possible condition is to just replace the
local rule $\lrule$ by a completely positive map, and insist on
commutativity as before. This does define a global evolution step
\cite{RiWe,Tak}. In view of the reversible case this would seem
like a good definition. However, it is again not clear whether the
composition of two such transformations will again be of the same
kind. Nor is it clear how to describe the class of transformations
obtained by several such ``simultaneous independent update''
steps.

It turns out that the analysis of Theorem~\ref{mainthm} can partly
be repeated in the irreversible case, thereby reducing the
possibilities somewhat. This line will be pursued elsewhere.

\subsection{Wolfram}
In a recent thick book \cite{Wolfram} S. Wolfram has argued that
the that the universe might be a big cellular automaton following
simple rules (see also \cite{Zuse}). Wolfram uses only classical
structures, expressing the belief, however, that quantum
structures might emerge from classical rules generating sufficient
complexity. Since Einstein failed with a program like that (he
thought of overdetermined non-linear field equations, rather than
CAs) it would be nice to see the details worked out.

\subsection{Quantum random walks}
The term ``Quantum cellular automaton'' has sometimes been used
\cite{Zeilinger,Iwo,Meyer,konno} for a unitary evolution of a
particle on a discretized space. The total Hilbert space of such a
system is $\ell^2(\Ir^s)$, the space of square summable complex
functions on the lattice, i.e., the direct sum rather than the
tensor product of the one-cell spaces. The classical analogue of
such a system is a single classical particle moving on a lattice,
e.g., in a random walk. Therefore much better terminology to call
these quantum systems {\it quantum random walks}, rather than
cellular automata. Such systems have been proposed for purposes of
quantum computation, in particular for search problems on graphs
(see \cite{Kempe} for a review).

Quantum random walks require localization properties not unlike
those of QCAs. To see the connection it is interesting to consider
the connection between classical random walks and CAs: A random
walk can be seen as a special CA, with each cell either empty or
occupied, started in a configuration with exactly one occupied
cell. Of course, the overall CA dynamics should respect this
constraint. Because the CA rule is local we can then also put
arbitrarily many particles on the lattice, and the dynamics is
well-defined by the random walk, as long as the particles do not
collide. What happens on collision is a piece of information which
must be supplied in order to make a random walk into a CA, i.e.,
in order to pass from a random walk to a (possibly
``interacting'') {\it diffusion}.

Consider a QCA with a special ``empty'' state specified for the
single cell algebra. This means that we can define a global
quantity {\it particle number}, which ought to be conserved by the
QCA. Technically this is the infinite sum of 0's and 1's, and not
a well defined observable. However, we can consider this formal
sum as a lattice interaction generating the time evolution of
``gauge transformations'', acting as an infinite product of
unitaries, as in Section~\ref{sec:communitary}. For a QCA
commuting with such transformations, the ``one-particle'' Hilbert
space can be defined, and the QCA dynamics restricted to this
subspace is a unitary evolution of the random walk type. Note that
we have not assumed that the dimension $d$ of the one-site algebra
is $2$, i.e., occupied cells (``particles'') may have an internal
structure. This turns out to be necessary: a standard classical
random walk, with only the empty/occupied distinction and nearest
neighbor interaction cannot be reversible. Similarly, a unitary on
$\ell^2(\Ir^s)$ cannot be strictly local \cite{Zeilinger}, and we
do need internal states \cite{Meyer2}. If there is only one
particle, this is the same as saying that we have only the
empty/occupied distinction for the cells, but we have also a
``quantum coin'' which helps determining the steps.

The standard model of a quantum random walk \cite{Kempe} uses a
qubit coin, i.e., three states for the QCA. All random walks with
such a coin are parameterized by a single unitary
$2\times2$-matrix $U$: we can describe the internal states
(``chirality'' \cite{konno}) as ``go right'' and ``go left''. A
trivial, but globally well-defined evolution step is defined by
following these instructions. The general case arises by following
this with a sitewise unitary rotation by $U$, representing the
quantum coin flip.

Now the question arises: can we consider such random walks as the
one-particle component of a QCA allowing arbitrarily many
particles? Indeed this is possible, even in many ways, and it
would be very interesting to classify all possibilities, hence all
``interactions''. One possibility is to second quantize the random
walk, which leads to a Boson system allowing, at each site, an
arbitrary number of particles. It is clear from the formalism of
second quantization that all the locality properties required of a
QCA are then satisfied. Clearly, this is the non-interacting
option. If we insist on finitely many states per cell we introduce
some kind of hard core interaction. A trivial way of doing it with
four states is this: Consider two infinite qubit chains, called
the left moving and the right moving chain. The ``free'' time
evolution $S$ is indeed shifting the two chains separately
according to this description. Now at each lattice site we take
the right moving and the left moving one-site algebras together to
define a single cell of the QCA, with the four states labelled
`empty', `R', `L', and `RL'. We then select a unitary $U$ leaving
the empty and the doubly occupied state invariant, but shuffling
the `R' and `L' states as before. Clearly, the combination of the
free evolution $S$ with the sitewise application of $U$ defines a
QCA whose one-particle sector is the given random walk.

\subsection{Spacetime localized algebras}

A quantum field theory can be considered from two equivalent
points of view. On the one hand one can consider the fields at
some time fixed time as the basic dynamical variables, e.g., the
Cauchy data at time zero. On the other hand, for the discussion of
relativistic causality, it is convenient to consider as
fundamental the family of algebras $\AO{}$ associated with the
measurements in some space-time region $\mathcal O$ \cite{Haag}.
Similarly, one can look at a QCA from these two points of view,
the ``Cauchy data'' point of view being what we described so far.
For the space-time view consider the extended lattice $\Ir^{s+1}$
of space-time points $(x,t)$, with $x\in\Ir^s$, the same spatial
lattice as before, and $t\in\Ir$ a discrete time point.

The global C*-algebra will be the same as before, but we introduce
additional subalgebras, namely
\begin{equation}\label{Axt}
    \AA_{x,t}=\grule^{-t}(\AA_x)\;.
\end{equation}
Since $\grule$ commutes with lattice translations we can define a
joint set of space-time translations $\grule_{x,t}$, combining a
lattice translation by $x$ with $\grule^{-t}$. For any subset
$\OO\subset\Ir^{s+1}$ we define $\AO{}$ as the C*-algebra
generated by all the $\AA_{x,t}$ with $(x,t)\in\OO$. In order to
study the localization properties of of these algebras, let us say
that a sequence of lattice points
\begin{equation}\label{slpath}
    (x_0,t_0),\ (x_1,t_0+1),\ (x_2,t_0+2),\ldots,(x_N,t_N)
\end{equation}
is (forward) {\it timelike}, if $x_{k+1}-x_k\in\Nei$ for all $k$.
Then we define two regions $\OO_1,\OO_2\subset\Ir^{s+1}$ to be
{\it spacelike separated} (notation: $\OO_1\spacelike\OO_2$), if
no timelike curve passes through some point of $\OO_1$ and also
through some point of $\OO_2$. Moreover, we say that $\OO_1$ is
{\it causally dependent} on $\OO_2$ (notation:
$\OO_1\vartriangleleft\OO_2$), if every timelike curve which
passes through (any point of) $\OO_1$ also passes through $\OO_2$.

Then it is clear that $\OO_1\spacelike\OO_2$ implies that $\AO1$
and $\AO2$ commute elementwise. Moreover, if $\OO_2$ is a finite
set of lattice points $(x,t)$, all with the same $t$, and if
$\OO_1\vartriangleleft\OO_2$, then $\AO1\subset\AO2$.

Conversely, if we have a net of local algebras $\AO{}$, defined
for arbitrary finite subsets $\OO$ of the spacetime-lattice, if
the spacetime translations act by automorphism of the total
algebra such that $T_{(x,t)}(\AO{})={\mathfrak A}({\OO}+(x,t))$,
and that the above locality properties hold, then the time zero
algebras form a QCA in the sense of Section~\ref{sec:def}.

This way of viewing QCAs may indeed be useful for making the
connection to relativistic field theories, or for a detailed study
of the growth of localization regions.

\subsection{Non-commuting cells}
As is standard in quantum theory we described the cells of the
automaton as subsystems in the usual tensor product sense. In a
slightly more axiomatic style we could have postulated, that
arbitrary measurements in separate cells can be carried out
jointly, implying the commutation between the different cells.
While this is certainly very natural, more general commutation
rules between cells may be considered. An obvious example are
anti-commutation rules, i.e., we could think of a Fermi gas on a
lattice. This would, of course, change the requirements on local
rules, but it is quite clear how to adapt our definition to that
case.

Another very interesting deviation from commuting cells is studied
in \cite{Seiler}, in connection with non-commutative analogs of
2-surfaces with constant negative curvature, the sine-Gordon
equation, and 2-dimensional lattice systems in  magnetic field
(Hofstadter butterfly). Roughly speaking the structure
investigated has has a single variable, say a unitary $Q_x$, at
each cell ``$x$''. Non-commutativity comes in, because any pair of
neighboring unitaries forms a discrete Weyl system. At distances
$\geq2$ the variables commute. It is clear that (reversible)
dynamical rules can be described exactly along the lines of our
definition, as automorphisms respecting this algebraic structure,
and also the localization (up to a finite enlargement of
localization regions). Of course, this is not the place to explore
such structures, and we point again to the book \cite{Seiler} for
more material.

\appendix
\section{Finite Dimensional C*-algebras}\label{APPCstar}

Almost all the hard work in any textbook on C*-algebras goes into
aspects of the theory, which become entirely trivial for finite
dimensional C*-algebras. Of course, some algebras in this paper
are infinite dimensional, notably the quasi-local C*-algebra
describing the infinite system. But the key arguments use only the
finite dimensional structure. Therefore we will give in this
Appendix a quick summary of C*-algebra theory as it applies to the
finite dimensional case.

In order to make the paper more accessible to communities in which
algebraic terminology is less current (e.g., most theoretical
physicists, and classical computer scientists) we start out with
an extended glossary, in which the basic notions are defined and
some basic facts are noted. This will be followed by some
structure theorems which we need in the body of the paper. Of
course, all this cannot replace a serious textbook. We recommend
\cite{BraRo,Tak,Dix}.

\subsection{C*-Glossary}

The operations making up the abstract structure of C*-algebras are
inspired by those known from algebras of operators on a Hilbert
space. In fact, every algebra of operators on a finite dimensional
Hilbert space, which is also closed under taking adjoints,
satisfies the definition, and every abstract C*-algebra is
isomorphic to an operator algebra. The following list of relevant
operations and concepts may serve as a glossary. The algebra under
consideration is usually denoted by $\AA$.

\begin{itemize}
\item {\it Addition} and multiplication by {\it complex scalars}. This makes
$\AA$ a vector space over $\Cx$, which we assume to be finite
dimensional from now on.

\item {\it Multiplication}.  We denote the product by $AB\in\AA$,
when $A,B\in\AA$. The product is distributive and associative (but
not necessarily commutative). If the algebra is commutative or
``abelian'' the system under consideration is classical.

\item The {\it adjoint} or ``star operation'' denoted by $A^*\in\AA$,
when $A\in\AA$. This is conjugate linear (or ``antilinear''),
which means that $(\lambda A)^*=\overline{\lambda}A^*$, and
$(A+B)^*=A^*+B^*$. The adjoint satisfies $(AB)^*=B^*A^*$.
Physicists often write $A^\dagger$ for the adjoint.

\item The {\it norm}, which is a positive number $\norm{A}$
associated with each $A\in\AA$. With respect to the algebraic
structures, the norm satisfies $\norm{A+B}\leq\norm A+\norm B$,
$\norm{\lambda A}=|\lambda|\;\norm A$, $\norm{AB}\leq\norm A\norm
B$ and $\norm{A^*A}=\norm A^2$. $\norm A=0$ implies $A=0$. In
contrast to the general case, the norm is uniquely determined by
the algebraic structures.

\item An {\it identity}. In contrast to the general case, in finite
dimensional C*-algebras there is always a unique element $\idty$
satisfying $\idty A=A\idty=A$.
\item A {\it homomorphism} between C*-algebras is a map $\Phi:\AA\to\BB$
between C*-algebras, preserving the algebraic structures. That is,
$\Phi$ is linear, $\Phi(AB)=\Phi(A)\Phi(B)$, and
$\Phi(A^*)=\Phi(A)^*$. The latter property is sometimes emphasized
by speaking of *-homomorphims. Homomorphisms $\Phi:\AA\to\AA$ are
called {\it endomorphisms} of $\AA$, and if an inverse
homomorphism exists, $\Phi$ is called an {\it isomorphism} or an
{\it automorphism} (if $\AA=\BB$). For general homomorphisms,
$\Phi(\idty)$ is a projection in $\BB$. If $\Phi(\idty)=\idty$, we
call it {\it unital}.
\item An {\it ordering}. We write $A\geq0$, if there is a
$B\in\AA$ such that $A=B^*B$. The set of positive elements is a
convex cone in the set of hermitian ($A^*=A$) elements. In an
operator algebra, the positive elements are precisely those
hermitian ones with all eigenvalues non-negative.
\item The {\it center} of $\AA$ is the subalgebra $\ZZ(\AA)\subset\AA$ of
elements $Z$ such that $ZA=AZ$ for all $A\in\AA$.
\item A {\it state} on $\AA$ is a linear functional
$\omega:\AA\to\Cx$ which is positive (i.e.,  $A\geq0$ implies
$\omega(A)\geq0$) and normalized (i.e., $\omega(\idty)=1$).
\item A {\it trace} on $\AA$ is a positive linear functional $\tau$ such
that $\tau(AB)=\tau(BA)$.

\item a positive linear functional $\omega$ is called {\it
faithful} if $A\geq0$ and $\omega(A)=0$ imply $A=0$. Every finite
dimensional C*-algebra has a faithful trace. On an operator
algebra, the usual trace of operators (which we will denote by
$\tr$) is an example.
\item Given a faithful trace $\tau$, every positive linear
functional can be written as $\omega(A)=\tau(\rho A)$, for a
unique $\rho\in\AA$, $\rho\geq0$,  which is called the {\it
density operator} of $\omega$ with respect to $\tau$. $\omega$ is
also a trace iff $\rho\in\ZZ(\AA)$.

\item An element $P\in\AA$ is a called a {\it projection} if $P^*=P=P^2$.
It is called a {\it minimal projection} if for any projection $Q$,
$Q\leq P$ implies $Q=0$ or $Q=P$.

\item The {\it direct sum} $\AA=\bigoplus_\mu\AA_\mu$ of a finite collection of
finite dimensional C*-algebras $\AA_\mu$ is the vector space
direct sum, i.e., elements are tuples of components
$A_\mu\in\AA_\mu$, with componentwise algebraic operations. In
operator algebras each term in the sum corresponds to a diagonal
block in a block matrix decomposition.
\item The {\it tensor product} $\AA=\bigotimes_\mu\AA_\mu$ is the
vector space tensor product, with product and adjoint defined as
the unique linear (resp. conjugate linear) extensions of
$(\otimes_\mu A_\mu)(\otimes_\mu B_\mu)=\otimes_\mu (A_\mu B_\mu)$
and $(\otimes_\mu A_\mu)^*=\otimes_\mu A_\mu^*$. In operator
algebras one forms this product by taking first the tensor
products of the underlying Hilbert spaces, and taking the algebra
generated by all tensor product operators.
\end{itemize}

\noindent The first four items on this list are the definition of
C*-algebras. Note that order and unit are explicitly defined
in terms of the algebraic structure (linear operations,
multiplication and adjoint), and the norm is also defined
explicitly as
\begin{equation}\label{defnorm}
  \norm A=\inf\{\lambda>0| \exists_B\;A^*A+B^*B=\lambda^2\idty\}
\end{equation}
It might thus seem superfluous to list the norm among the defining
elements. In the infinite dimensional case it is needed, of
course, to formulate the topological completeness requirement (and
completeness is needed in turn to construct $B$ in
(\ref{defnorm})). However, even in the finite dimensional case the
implication $(\norm A=0)\Rightarrow (A=0)$ carries non-trivial
information by excluding the existence of nonzero elements $A_i$
such that $\sum_iA_i^*A_i=0$.

\subsection{C*-structure}

Consider a single Hermitian element $A\in\AA$. Then since $\AA$ is
finite dimensional, the powers $A^n$ must be linearly dependent,
i.e., there is a characteristic polynomial $p(A)=\sum_kc_kA^k=0$.
From this one readily constructs polynomials $p_\ell$ such that
$p_\ell(A)$ is a projection, and
\begin{equation}\label{babyspectral}
  A=\sum_\ell a_\ell p_\ell(A)
\end{equation}
where $a_\ell$ are the distinct roots of $p(a)=0$. This is called
the {\it spectral theorem}. It implies, in particular, that any
finite dimensional C*-algebra has many projections (which may fail
in infinite dimension).

This fact will be used in the following fundamental structure
theorem. Recall that by $\MM_n$ we denote the algebra of complex
$n\times n$ matrices.

\begin{proposition}\label{Csform}
Every finite dimensional C*-algebra $\AA$ is
characterized uniquely up to isomorphism by a finite sequence
$n_1\geq n_2\geq\cdots\geq1$ of numbers such that
\begin{equation}\label{isoCstar}
  \AA \cong\bigoplus_\mu \MM_{n_\mu}\;.
\end{equation}
\end{proposition}

The basic idea of the proof is to consider the center of $\AA$,
which is a finite dimensional abelian C*-algebra. The minimal
projections $z_\mu$ of the center decompose the algebra into a
direct sum $\AA=\bigoplus_\mu z_\mu\AA$, in which each of the
summands has trivial center. The building blocks $z_\mu\AA$ are
then seen to be isomorphic to full matrix algebras $\MM_{n_\mu}$,
where $n_\mu$ is the maximal number of mutually orthogonal
projections in $z_\mu\AA$.

\begin{proposition}\label{Cshom}
If $\Phi:\MM_d\to \bigoplus_\mu\MM_{n(\mu)}$ is a *-homomorphism
such that $\Phi(\idty)=\idty$, each $n(\mu)$ has to be divisible
by $d$.
\end{proposition}

By considering the composition of the given homomorphism with the
projection onto one summand, which is also a homomorphism, we can
consider the case of a single summand. Thus $\Phi$ becomes a
representation of $\MM_d$ on a Hilbert space of dimension
$n(\mu)$, which can be decomposed into irreducible
representations. It is a basic property of $\MM_d$, however, that
all its irreducible representations are unitarily equivalent to
the defining representation on $\Cx^d$, so $n(\mu)$ must be $d$
times the multiplicity (number of isomorphic irreducible summands)
of this representation.

\begin{proposition}\label{Cscom}
If $\AA\cong\bigoplus_\mu \MM_{n(\mu)}$ and $\BB\cong\bigoplus_\nu
\MM_{m(\nu)}$ are commuting subalgebras of $\BB(\HH)$, the algebra
$\AA\BB$ is also decomposed into direct summands, each of which
arises by multiplying a summand form each of the algebras. These
occur with  integer multiplicities $r_{\mu\nu}\geq0$ such that
\begin{equation}\label{multip}
 \sum_{\mu\nu}r_{\mu\nu}n(\mu)m(\nu)
  =\dim\HH\;.
\end{equation}
\end{proposition}

Let $A_\mu\in\MM_{n(\mu)}\subset\AA$ and
$B_\nu\in\MM_{m(\nu)}\subset\BB$ be elements from the respective
blocks. Then $A_\mu\otimes B_\nu\mapsto A_\mu B_\nu$ is a
representation of $\MM_{n(\mu)\cdot
m(\nu)}\cong\MM_{n(\mu)}\otimes\MM_{m(\nu)}$ on $\HH$, which may
however be zero since we cannot guarantee that it preserves the
identity. The rest of the argument is as for the previous
proposition.

\section{Acknowledgements}
The main lines of this paper were conceived in March 2003, in
discussions at the Centro Ettore Majorana in Erice , whose
hospitality we acknowledge. We also benefited from discussions
with I. Cirac, M. Wilkens, and C.H. Bennett.


\begin{thebibliography}{99}

\bibitem{Feynman}R. Feynman, Simulating physics with computers,
{\it Int. J. Theor. Phys. \bf21} (1982) 467-488, Reprinted in
A.J.G. Hey (ed.), {\it Feynman and Computation}, Perseus Books
1999.

\bibitem{optlat}O. Mandel, M. Greiner, A. Widera, T. Rom, T.W. H\"ansch,
  and I. Bloch,
  \jtitle{Coherent transport of neutral atoms in spin-dependent optical
     lattice potentials,}
  {\it Phys. Rev. Lett. \bf91}, 010407 (2003)

\bibitem{mictrap}R. Dumke, M. Volk, T. Muether, F.B.J.
   Buchkremer, G. Birkl, and W. Ertmer,
   \jtitle{ Microoptical Realization of Arrays of Selectively Addressable
    Dipole Traps: A Scalable Configuration for Quantum Computation
    with Atomic Qubits,}
   {\it Phys. Rev. Lett. \bf89}, 097903 (2002) and quant-ph/0110140

\bibitem{QdCA} Further away are proposals based on arrays of
quantum dots, which are also published under the heading ``quantum
cellular automata''. These are ideas for new hardware for {\it
classical} computing, possibly replacing CMOS technology. In order
to avoid confusion with the ideas in which quantum coherence plays
a key role (as it does in our paper), many authors from that
community are now using the more precise term ``quantum {\it dot}
cellular automata''. For an overview see the home page of the
Notre Dame group (www.nd.edu/\~{}qcahome), or: P.D. Tougaw, C.S.
Lent, \jtitle{Logical devices implemented using quantum cellular
automata,} {\it J.Appl.Phys. \bf75} (1994) 1818.

\bibitem{Benjamin} S. C. Benjamin,
  \jtitle{Schemes for parallel quantum computation without local
  control of qubits,}
  {\it Phys. Rev. A \bf61} 020301 (2000)

\bibitem{Watrous}J. Watrous: On one-dimensional quantum cellular automata.
In Proceedings of the 36th Annual Symposium on Foundations of
Computer Science, 1995, pp.~528--537.

\bibitem{vanDam}W. van Dam: Quantum cellular automata, Master
Thesis, Computer Science Nijmegen, Summer 1996

\bibitem{Gruska}J. Gruska: {\it Quantum Computing}, (McGraw-Hill,
Cambridge 1999). QCAs are treated in Section 4.3.

\bibitem{santa}C. D\"urr and M. Santha: A decision procedure for
unitary linear quantum cellular automata,
quant-ph/9604007;\newline
 C. D\"urr, H. L\^eTanh and M. Santha,
 \jtitle{A decision procedure for
           well-formed linear quantum cellular automata,}
 {\it Rand. Struct. Algorithms \bf11}, 381-394 (1997)
 and cs.DS/9906024

\bibitem{Brennen}G. K. Brennen and J. E. Williams,
   \jtitle{Entanglement dynamics in 1D quantum cellular automata, }
   quant-ph/0306056

\bibitem{BraRo}O. Bratteli and D. Robinson:
{\it Operator algebras and quantum statistical mechanics}, vol. I
(Springer 1979)

\bibitem{Paschen}A C*-algebraic framework was also applied to QCAs
in a recent thesis. However, it was applied to the state side,
rather than the observables, making locality properties harder to
see. K. Paschen: \"Uber Reversibilit\"at, Nicht-Determiniertheit
und Quantenrechnen in Zellularautomaten, Dissertation in
Informatik (PhD Thesis in Computer Science), Karlsruhe 2002

\bibitem{Haag}R. Haag: {\it Local quantum physics}, Springer 1996

\bibitem{RiWe}S. Richter and R.F. Werner,
  \jtitle{Ergodicity of quantum cellular automata,}
  {\it J. Stat. Phys. \bf82} (1996) 963-998 and cond-mat/9504001

\bibitem{Kari1}J. Kari,
  \jtitle{On the circuit depth of structurally reversible cellular automata,}
   {\it Fund.Inform. \bf34} (2003) 1–-15

\bibitem{otherlat}The basic concepts in this paragraph work in any lattice
structure. In fact, they do not even require translation
invariance and could be formulated for possibly different spins
(given by possibly infinite dimensional C*-algebras) localized on
the nodes of a finite or infinite graph.

\bibitem{onewaycomp}R. Raussendorf, D. E. Browne and H.-J. Briegel,
   \jtitle{The one-way quantum computer - a non-network model of quantum
           computation,}
   {\it J. Mod. Opt \bf 49}, 1299 (2002).

\bibitem{Rich}D. Richardson,
   \jtitle{Tesselation with local transformations,}
   {\it J.Comp.Syst.Sci. \bf6} (1972) 373--388

\bibitem{classicalW}What appears here as a pathology is allowed in
the (periodic boundary version) of Watrous QCAs, i.e., the
Wrapping Lemma fails for that structure. A more systematic
study of the possibility shown by our example was carried out in\\
S. Inokuchi, Y. Mizoguchi,
 \jtitle{Generalized partitioned quantum
cellular automata and quantumization of classical CA,}
quant-ph/0312102.

\bibitem{Lloyd}S. Lloyd,
    \jtitle{A potentially realizable quantum computer,}
    {\it Science \bf261}, 1569--1571 (1993)

\bibitem{cliff} We follow this terminology, although it is not clear what
Clifford had to with this. In field theory these transformations
would be called ``quasi-free'', or ``Bogolyubov automorphisms'',
in phase space quantum mechanics ``metaplectic transformations''.

\bibitem{OurCliff}D. Schlingemann, R.F. Werner, The structure of
Clifford quantum cellular automata, in preparation.

\bibitem{MarTo}T. Toffoli and M. Margolus,
 \jtitle{Invertible Cellular automata: a review,}
 {\it Physica D\bf 45}(1990) 229-253

\bibitem{Zanardi}The support algebra of a single element, an
interaction Hamiltonian was also introduced under the name
``interaction algebra'' by P.Zanardi: Stabilization of quantum
information: a unified dynamical-algebraic approach,
quant-ph/0203008

\bibitem{Wootters} K.S. Gibbons, M.J. Hoffman, W.K. Wootters,
  \jtitle{Discrete phase space based on finite fields,}
  quant-ph/0401155

\bibitem{Paz}J. P. Paz: Discrete Wigner functions and the phase space representation of
quantum teleportation, quant-ph/0204150

\bibitem{Aoun}B. Aoun, M. Tarifi, Quantum cellular automata,
quant-ph/0401123.

\bibitem{Tak}M. Takesaki, {\it Theory of operator algebras, I},
Springer 1979

\bibitem{Wolfram} S. Wolfram, {\it A new kind of science},
(Self-published, Wolfram Media Inc. 2002)

\bibitem{Zuse} K. Zuse, {\it Rechnender Raum},
Schriften zur Datenverarbeitung, Band 1, Vieweg,
 Braunschweig 1969.

\bibitem{Zeilinger}G. Gr\"ossing and A. Zeilinger,
  \jtitle{Quantum cellular automata,}
  {\it Complex Systems \bf2}, 197--208 (1988)

\bibitem{Iwo}I. Bialynicki-Birula,
  \jtitle{Weyl, Dirac, and Maxwell equations on a lattice as unitary cellular automata, }
  {\it Phys. Rev. D \bf49}, 6920--6927 (1994)

\bibitem{Meyer}D. A. Meyer,
  \jtitle{From quantum cellular automata to quantum lattice gases, }
  {\it J. Stat. Phys. \bf85},551--574 (1996)

\bibitem{konno}N. Konno, K. Mitsuda, T. Soshi, H.J. Yoo,
  \jtitle{Quantum walks and reversible cellular automata},
  quant-ph/0403107

\bibitem{Kempe}J. Kempe,
 \jtitle{Quantum random walks: an introductory overview,}
 {\it Contemp.Phys. \bf44} (2003) 307 -– 327

\bibitem{Meyer2}D. A. Meyer: Unitarity in one dimensional nonlinear
quantum cellular automata, quant-ph/9605023;
 From quantum cellular automata to quantum lattice gases, {\it
 J.Stat.Phys. \bf85} (1996) 551--574

\bibitem{Seiler}A.I. Bobenko, R. Seiler (eds.),
 {it Discrete integrable geometry and physics},
 Clarendon Press, Oxford 1999. The book has several articles
 connecting to the strucutre mentionend in the text. It is best to
 pick up the pointers in the introduction by the editors.

\bibitem{Dix}J. Dixmier, {\it C*-algebras}, North Holland 1977






\end{thebibliography}
\end{document}